\documentclass[%
reprint,
superscriptaddress,
showpacs,
showkeys,
 amsmath,amssymb,
aps,
pra,
longbibliography,
floatfix
]{revtex4-1}

\usepackage{graphicx}
\usepackage{dcolumn}
\usepackage{bm}
\usepackage[english]{babel}

\usepackage[utf8]{inputenc}
\usepackage[T1]{fontenc}

\usepackage{natbib}
\usepackage{letltxmacro}

\LetLtxMacro{\ORIGselectlanguage}{\selectlanguage}
\makeatletter
\DeclareRobustCommand{\selectlanguage}[1]{%
  \@ifundefined{alias@\string#1}
    {\ORIGselectlanguage{#1}}
    {\begingroup\edef\x{\endgroup
       \noexpand\ORIGselectlanguage{\@nameuse{alias@#1}}}\x}%
}
\newcommand{\definelanguagealias}[2]{%
  \@namedef{alias@#1}{#2}%
}
\makeatother

\definelanguagealias{en}{english}

\usepackage{color}
\usepackage[unicode]{hyperref}
\hypersetup{
   unicode=true,          
   plainpages=false,
   colorlinks=true,       
   citecolor=blue,        
}

\usepackage{url}

\newcommand{\rref}[1]{Eq.\ (\ref{#1})}
\newcommand{\rrefsa}[1]{Eqs.\ (\ref{#1})}
\newcommand{\rrefsb}[1]{(\ref{#1})}

\newcommand{\km}{k_\mathrm{max}}
\newcommand{\Ex}{{E}}
\newcommand{\px}{{\Phi}}
\newcommand{\pa}{{\Phi_\mathrm{a}}}
\newcommand{\pL}{{\Phi^\mathrm{L}_\mathrm{a}}}
\newcommand{\Ea}{{E_\mathrm{a}}}

\begin{document}


\title{Accelerating the convergence of exact diagonalization with the transcorrelated method: 
Quantum gas in one dimension with contact interactions}

\date{\today}

\author{P\'eter Jeszenszki}
 \email{jeszenszki.peter@gmail.com}
\affiliation{Dodd-Walls Centre for Photonics and Quantum Technology, Dunedin 9054, New Zealand}
\affiliation{New Zealand Institute for Advanced Study, and Centre for Theoretical Chemistry and Physics, Massey University, Private Bag 102904 North Shore, Auckland 0745, New Zealand
}%
\affiliation{
 Max Planck Institute for Solid State Research, Heisenbergstra{\ss}e 1,
70569 Stuttgart, Germany
}%


\author{Hongjun Luo}
\email{H.Luo@fkf.mpg.de}
\affiliation{
 Max Planck Institute for Solid State Research, Heisenbergstra{\ss}e 1,
70569 Stuttgart, Germany
}%
\author{Ali Alavi}
\email{A.Alavi@fkf.mpg.de}
\affiliation{
 Max Planck Institute for Solid State Research, Heisenbergstra{\ss}e 1,
70569 Stuttgart, Germany
}%
 \affiliation{Department of Chemistry, University of Cambridge,
Lensfield Road, Cambridge, CB2 1EW, United Kingdom}
\author{Joachim Brand}%
 \email{J.Brand@massey.ac.nz}
\affiliation{Dodd-Walls Centre for Photonics and Quantum Technology, Dunedin 9054, New Zealand}
\affiliation{New Zealand Institute for Advanced Study, and Centre for Theoretical Chemistry and Physics, Massey University, Private Bag 102904 North Shore, Auckland 0745, New Zealand
}%
\affiliation{
 Max Planck Institute for Solid State Research, Heisenbergstra{\ss}e 1,
70569 Stuttgart, Germany
}%
\begin{abstract} 
Exact diagonalization expansions of Bose or Fermi gases with contact interactions converge very slowly due to a nonanalytic cusp in the wave function.
Here we develop a transcorrelated approach where the cusp is treated exactly and folded into the many-body Hamiltonian with a similarity transformation that removes the leading-order singularity. The resulting  transcorrelated Hamiltonian is not Hermitian but can be treated numerically with a standard projection approach. The smoothness of the wave function improves by at least one order and thus the convergence rate for the ground-state energy improves.
By numerical investigation of a one-dimensional gas of spin-$\frac{1}{2}$ fermions we find the error in the transcorrelated energy to scale as  $M^{-3}$ with a single-particle basis of $M$ plane waves compared to $M^{-1}$ for the expansion of the original Hamiltonian and $M^{-2}$ using conventional lattice renormalization.
\end{abstract}

\maketitle


\section{\label{sec:Introduction}Introduction}
In recent years there has been increasing interest in the experimental realization of  strongly correlated quantum gases 
with bosonic or fermionic ultracold atoms \cite{Kinoshita2006,Haller2009,Liao2010,hueck_two-dimensional_2018,van_wyk_superfluid_2018}. 
Their theoretical description is difficult  \cite{Cazalilla2011,giorgini_theory_2008}
and efficient numerical 
methods are required to describe the system reliably and accurately. One straightforward approach is to  diagonalize the Hamiltonian in a Fock basis, i.e., a finite basis of appropriately symmetrized products of single-particle wave functions
\cite{Deuretzbacher2007,damico_pairing_2015,pecak_experimentally_2017,grining_many_2015,sowinski_pairing_2015,rammelmuller_evolution_2017,garcia-march_sharp_2013,garcia-march_quantum_2014}. 
Besides the energy, this approach also provides convenient access to the full wave function from which all system properties can be computed.
With the recently 
developed Full Configuration Interaction Quantum Monte Carlo method \cite{booth_fermion_2009} it has become possible to solve much larger problems than with conventional, deterministic approaches \cite{thomas_accurate_2015,booth_towards_2012,li_manni_combining_2016}. 
Nevertheless, the exact diagonalization in Fock space
is computationally  expensive as the size of the many-particle Fock basis grows combinatorially with the number of particles and the size of the single-particle basis. Moreover,  the convergence towards exact results with increasing the size of the single-particle basis is painfully slow for short-range interacting ultracold atoms. Specifically, for a one-dimensional model with $\delta$ function interactions, which models ultracold atoms in a tightly confining trap \cite{olshanii_atomic_1998}, the energy converges to the exact result with an error that scales as $M^{-1}$ with a basis set of $M$ plane waves, or $M^{-1/2}$ with a basis set of harmonic-oscillator eigenfunctions  \cite{grining_many_2015}.
The reason for this slow convergence is the fact that the short-range interaction induces a cusp into the many-body wave functions at particle coalescence, i.e., whenever two particles meet \cite{lieb_exact_1963,olshanii_atomic_1998}. Mathematically, the wave function belongs to the differentiability class $C^0$, i.e., it is continuous but its first derivative is discontinuous. Approximating such a shape of the wave function by linear combinations of products of smooth single-particle functions is highly inefficient.

One possible way to improve energy estimates in a finite basis set is to renormalize the parameters of the Hamiltonian \cite{castin_simple_2004,ernst_simulating_2011,Werner2012}. This approach is closely related to the concept of a running coupling constant in quantum field theory, where the coupling constant depends on a momentum cutoff \cite{Wen2004}.
In two and three-dimensional systems with contact interaction it is necessary to renormalize the interaction constant with the basis size (momentum cutoff) in order to avoid divergences. While this is not necessary in one dimension, adjusting the interaction strength can still improve convergence properties \cite{ernst_simulating_2011}. The simplest possibility is to adjust the interaction constant such as to yield the exact value of the ground-state energy for two interacting particles from a calculation in the truncated basis. We are not aware whether the improved convergence rate of a many-body calculation has been determined before, but in Sec.~\ref{sec:Numcomp} we report numerical results for three and six fermions that indicate that the convergence rate of the energy error improves by one order from $M^{-1}$ to $M^{-2}$.
Renormalizing the coupling strength based on an exactly solvable limit of the many-body problem was proposed in Ref.~\cite{ernst_simulating_2011} and
the possibility of adjusting the dispersion relation of the kinetic energy part of the Hamiltonian in order to improve the convergence rate was discussed in Refs.~\cite{Carlson2011,Werner2012} for two- and three-dimensional Fermi gases.
In a similar spirit as the renormalization of the coupling constant, a more elaborate effective Hamiltonian approach has been used to speed up the convergence of few-particle calculations in harmonic trapping potentials \cite{lindgren_fermionization_2014,dehkharghani_quantum_2015}. Other approaches to optimize the finite-basis representation of the wave function have suggested modifying the cutoff procedure for the exact diagonalization procedure \cite{marcin_plodzien_numerically_2018}, or scaling the single-particle basis function \cite{Koscik2018}.


A different way to improve the convergence properties of a basis set expansion is to use basis functions that explicitly depend on the 
interparticle distances instead of a Fock basis \cite{von_stecher_correlated_2009,mitroy_theory_2013,brouzos_two-component_2013,przemyslaw_koscik_variational_2018}. However,
the cost of the determination of the matrix elements exponentially increases with the number of  particles restricting the applicability of the method
to the few-particle regime. 

The route that we follow in this work is to introduce a Jastrow factor $e^{\tau}$  \cite{jastrow_many-body_1955} in order to capture the short-range behavior of the exact $N$-particle wave function $\Psi$ 
\begin{align}
 \Psi(x_1, x_2, \dots, x_N )  &= e^{\tau}  \Phi(x_1, x_2, \dots , x_N) \ , \label{Jastrow}\\
 \tau &= \sum_{k<l}^N u( x_k - x_l) \ , \label{ansatz} 
\end{align}
where the correlation factor $\tau$ depends on the pairwise separation distances of particles.
The function
$u(x)$ is designed to describe important two-particle correlations of $\Psi$ while the  function $\Phi$ is much smoother.
%
Jastrow factors as in Eq.~\eqref{Jastrow} are  
frequently used in  Variational and
Diffusion Quantum Monte Carlo approaches  for ultracold atoms \cite{casula_quantum_2008,schonenberg_effective-range_2017,bertaina_bcs-bec_2011,pessoa_monte_2015,parisi_quantum_2017}, {\it ab initio} nuclear physics \cite{carlson_quantum_2015}, the electronic structure of atoms, molecules \cite{lipkowitz_quantum_2007,lester_quantum_2009}, and solid-state materials \cite{foulkes_quantum_2001,kolorenc_applications_2011}.

In this paper we follow the  transcorrelated approach \cite{boys_determination_1969}, where the Jastrow factor is folded into the Hamiltonian. Starting from the stationary Schr\"odinger equation
\begin{align} \label{eq:stationarySE}
\hat{H} \Psi = E \Psi ,
\end{align}
and using Eq.~\eqref{Jastrow} we obtain the transcorrelated  Schr\"o\-din\-ger equation 
\begin{align}
\underbrace{e^{-\tau} \hat{H} e^{\tau}}_{\tilde{H}} \Phi  = E  \Phi  \ , \label{effectiveScheq}
\end{align}
where the transcorrelated Hamiltonian $\tilde{H} = e^{-\tau} \hat{H} e^{\tau}$ is related to the original Hamiltonian $\hat{H}$ by a similarity transformation and thus shares the same eigenvalue spectrum. The transcorrelated method has already been widely used for computations of atomic, molecular
\cite{boys_determination_1969,filippi_multiconfiguration_1996,hino_application_2002,luo_variational_2010} and solid-state properties \cite{ochi_optimization_2014,ochi_iterative_2016}, where typically the emphasis has been on finding an optimized correlation factor $\tau$, while $\Phi$ is taken as a simple reference function with the correct particle exchange symmetry, e.g., a Slater determinant. Here, we  follow the idea of Ref.~\cite{luo_combining_2018} where the function
$u(x)$ is designed to exactly reproduce the singular short-range behavior of the exact wave function $\Psi$, while the transcorrelated function $\Phi$ is expanded in a Fock basis.
The transcorrelated Schr\"odinger equation 
 \rrefsb{effectiveScheq} is then solved as an exact diagonalization problem. 
Reference \cite{luo_combining_2018} demonstrated that an improved convergence rate and highly accurate energies for the homogeneous Coulomb gas could be achieved with this approach.

In the following, we concentrate on a one-dimensional quantum gas of bosons or fermions with contact interactions. By constructing a correlation factor with an appropriate cusp, we show that the smoothness of the transcorrelated wave function is improved by at least one order, i.e.~from $C^0$ to $C^1$ where the first derivative is continuous. For spinless bosons where the wave function is symmetric under pairwise particle exchange, the transcorrelated wave function even improves further to $C^2$, i.e.\ the second derivative is continuous as well. 
The explicit and exact form of the transcorrelated Hamiltonian is derived in real space, and in momentum space in second quantization. Three-particle interactions occur as a consequence of the similarity transformation. While a useful approximation is developed that only requires evaluation of effective two-particle terms, it is also shown that convergence to exact results for the energy can be achieved with a purely one- and two-body effective Hamiltonian if the correlation factor is appropriately adjusted with the basis set size.
While the transcorrelated approach developed here is not restricted to ground states or the absence of trapping potentials, we specifically consider
the cases of the homogeneous gas of spin-$\frac{1}{2}$ fermions and spinless bosons  and compare with  exact solutions based on the Bethe ansatz \cite{lieb_exact_1963,yang_exact_1967,gaudin_systeme_1967}
\footnote{Exact solutions are also available for inhomogeneous systems in the strongly interacting limit, which could be used for further benchmarking in future work \cite{girardeau_relationship_1960,deuretzbacher_exact_2008,deuretzbacher_quantum_2014,volosniev_strongly_2014,levinsen_strong-coupling_2015}.}.

This allows for easy benchmarking of our numerical results, which are presented for the spin-$\frac{1}{2}$ Fermi gas in a plane-wave basis. We find that the convergence rate of the energy error is improved by the transcorrelated method from $M^{-1}$ to $M^{-3}$ in a basis of $M$ single-particle functions.

This paper is organized as follows. After introducing the original Hamiltonian with $\delta$ interactions and discussing the wave function cusp in Sec.~\ref{sec:cusp},  we construct an appropriate correlation factor in Sec.~\ref{sec:correlationfactor} before deriving the explicit form of the transcorrelated Hamiltonian in real space in Sec.~\ref{sec:explicit}. In Sec.~\ref{sec:smoothness} we show that the correlation factor improves the smoothness of the wave function by one or two orders before providing an analytical estimate for the convergence rate of the energy in Sec.~\ref{subsec:Convergence rate}. Discussing the momentum space form of the transcorrelated Hamiltonian in second quantization and a convenient approximation for the three-body term in Sec.~\ref{sec:secondquant} concludes the theory part \ref{sec:Theory}. Numerical results for spin-$\frac{1}{2}$ fermions are presented in Sec.~\ref{sec:Numcomp}. After introducing the methods used in Sec.~\ref{sec:methods}, we present calculations of the energy error for two particles in Sec.~\ref{sec:two}, for three particles in Sec.\ \ref{sec:threeparticles}, and for six particles in Sec.~\ref{sec:sixparticles} before concluding in Sec.~\ref{sec:Conclusion}.


\section{\label{sec:Theory}Theory}

\subsection{One-dimensional quantum gas with contact interaction: cusp of the wave function}\label{sec:cusp}

We consider a gas of $N$ quantum particles of mass $m$ in one spatial dimension. Either a single or several spin flavors of bosons or fermions may be present.
The particles interact
with a contact (zero-range) interaction,  which can be represented by a Dirac delta potential in the
Hamiltonian 
\begin{eqnarray}
\hat{H} &=& -\frac{\hbar^2}{2m}\sum_{i=1}^N\left[ \frac{\partial^2}{\partial x^2_i} +
V(x_i)\right] +
g \sum_{i<j} \delta \left(x_i-x_j \right) , \label{scheq}
\end{eqnarray} 
where $x_i$ is the spatial coordinate of the $i$th particle and $V(x)$ a smooth external potential.
The external trapping potential $V(x)$ does not change the singlular properties of the wave function, which are dominated by the contact interaction term. For this reason, we will omit the potential $V(x)$ for many examples, which allows the comparison to exact solutions obtained by the Bethe ansatz, e.g.~for spinless bosons \cite{lieb_exact_1963} and spin-$\frac{1}{2}$ fermions \cite{yang_exact_1967,gaudin_systeme_1967}. Our transcorrelated approach for improving the smoothness of the wave function, however, is not restricted to the homogeneous system and the asymptotic convergence rates that we report in this work are not affected by smooth external potentials.

The Hamiltonian \eqref{scheq} can be realized with ultracold atoms in a tightly confining wave-guide-like trapping potential \cite{olshanii_atomic_1998}.
The potential strength $g$ can be expressed through a one-dimensional scattering length $a$ as
\begin{eqnarray}
g=- \frac{2\hbar^2}{ma} \ . \label{potstrengthtoscatt}
\end{eqnarray}
The contact interaction can be also expressed as boundary condition for the wave function at coalescence, i.e.\ when two particles meet \cite{lieb_exact_1963}. 
Note that fermions only feel the presence of the contact interaction term between different spin flavors due to the Pauli exclusion principle.
Near the coalescence point the wave function takes the form 
(in analogy to Refs.~\cite{bethe_quantum_1935,olshanii_short-distance_2003,Werner2012})
\begin{widetext}
\begin{align}
& \Psi(x_1, x_2, \dots ) \stackrel{x_{ij} \rightarrow 0}{=} 
\left( a -|x_{ij}| \right) 
A_{ij}(X_{ij}, x_1, \dots, x_{i-1},x_{i+1},  \dots, x_{j-1 },x_{j+1 }, \dots )  + \mathcal{O}(x_{ij}) , \label{BPOneD}
\end{align}
\end{widetext}
where $x_{ij}=x_i-x_j$ is the relative and $X_{ij}=(x_i+x_j)/2$ is the center-of-mass coordinate relating to the pair with the  $i$th and $j$th particles.
The $\mathcal{O}(x_{ij})$ term is regular at coalescence. 
For fermions the  function $A_{ij}(X_{ij}, x_1, \dots)$  is equal to zero if
both particles $i$ and $j$ belong to the same spin flavor due to the Pauli exclusion principle, which also means that they do not feel the contact interaction. The term $a -|x_{ij}|$ in \rref{BPOneD} describes a cusp in the wave function with a
discontinuity in the first derivative and a singularity in the second derivative. 
The wave function thus belongs to the differentiability class $C^0$.
The cusp in the wave function further has the consequence that the  Fourier transform of the wave function to momentum space has $k^{-2}$ tails for large $k$ and, thus, the single-particle momentum distribution falls off as $k^{-4}$, as is well known for quantum gases in one dimension with contact interactions \cite{olshanii_short-distance_2003}.   


\subsection{Correlation factor for 1D system with contact interaction}\label{sec:correlationfactor}

In this work we follow a similar procedure to Ref.\ \cite{luo_combining_2018} and design the correlation factor $\tau$ of Eq.\ \rrefsb{ansatz} such that the boundary condition \rrefsb{BPOneD} is satisfied automatically. To this end it is sufficient to require  the function $u(x)$ to have the form
\begin{eqnarray}u(x) \stackrel{x \rightarrow 0}{=} u(0) -\frac{1}{a}|x| +{\mathcal O}\left( x^2 \right) \ . \label{Taylorexpu}
\end{eqnarray}
This restriction is enough to obtain the correct boundary condition, which can be seen by substituting \rref{Taylorexpu}  into the Jastrow factor 
\begin{align}
e^\tau = & \ \left(\prod\limits_{k\ne i , l \ne j} e^{u(x_{k l})} \right)e^{u(0)}\underbrace{e^{ -\frac{1}{a} |x_{ij}| + {\mathcal O}\left( x_{ij}^2 \right)}}_{1 - \frac{1}{a} |x_{ij}| + \mathcal O \left( x_{ij}^2 \right) } \ .
\label{JastrowTaylor}
\end{align} 

It is convenient to define the correlation factor in momentum space
to have the correct large-$k$ dependence and a simple cutoff for small $k$
\begin{equation}
\tilde{u}(k) = 
\begin{cases}
\frac{2}{ak^2} & \text{if } |k| \ge k_c \ , \\
0             & \text{if } |k| < k_c \ ,
\end{cases}
\label{u1d}
\end{equation}
where the parameter $k_c$  sets an inverse length scale. The advantages of choosing this specific form of the correlation factor will become fully clear in the following sections. An important feature is the free parameter $k_c$, which controls the size of the correlation factor and becomes important for suppressing three-particle contributions in the transcorrelated Hamiltonian.
The function $u(x)$
can be obtained by the inverse Fourier transform $u(x) = (2\pi)^{-1}\int \exp(-ikx) \tilde{u}(k) \mbox{d}k$, as
\begin{align}
u(x)= \frac{2}{a\pi}\left(\frac{\cos \left( k_c x\right)}{k_c} +  x \, \mbox{Si} \left(k_cx \right) - \frac{\pi}{2}|x|\right) ,
\label{FTu0OneD}
\end{align}
where $\mbox{Si}(x)$ is the sine integral function \cite{eric_w._weisstein_sine_nodate}. The function $u(x)$ is found to be smooth except at the origin. 
Considering the case when $x$ is close to zero we obtain the
expression
\begin{eqnarray}
u(x) = \frac{2}{ak_c \pi}-\frac{1}{a} |x|+{\mathcal O}\left( x^2 \right) \ , \label{uOneDTaylor}
\end{eqnarray}
which satisfies the condition \rrefsb{Taylorexpu}.

\begin{figure}
\includegraphics[scale=0.6]{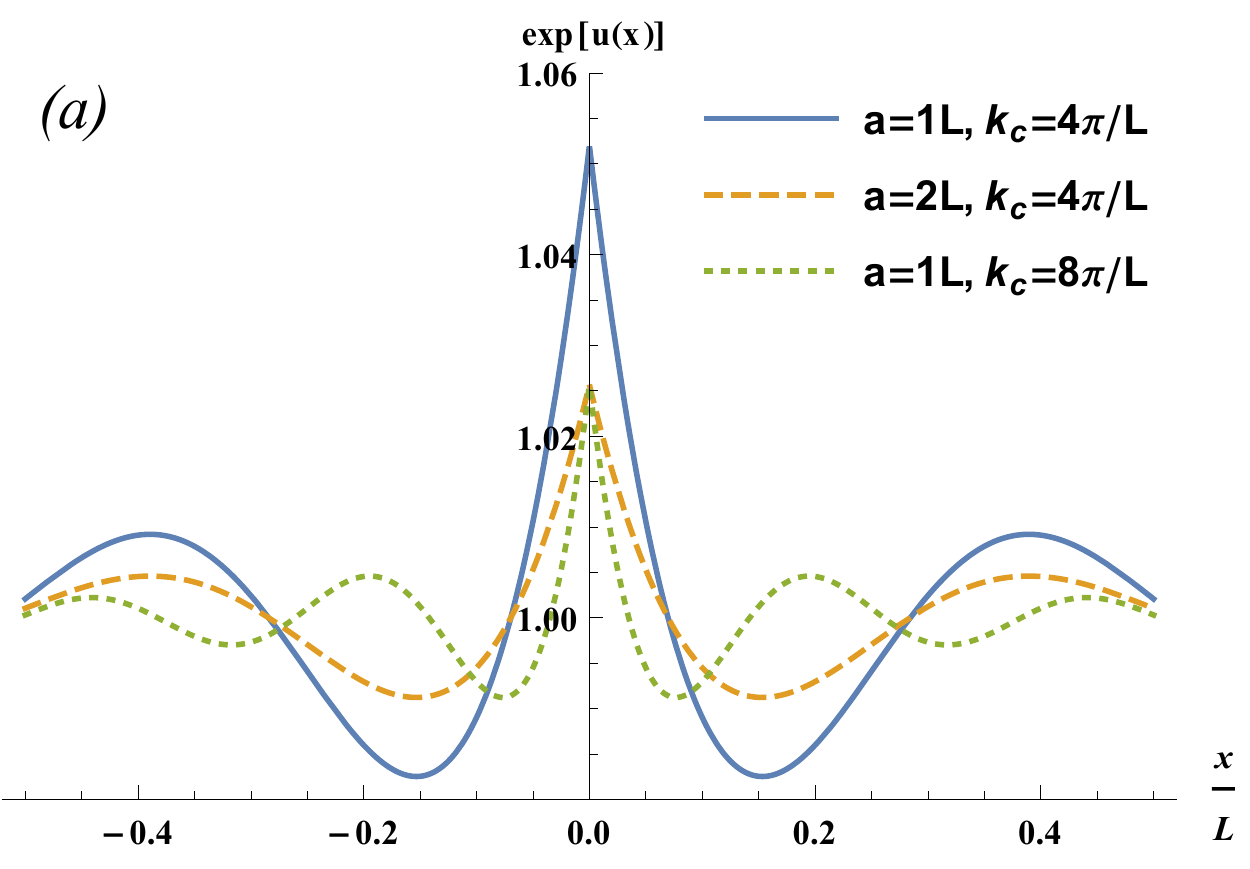}\\
\includegraphics[scale=0.6]{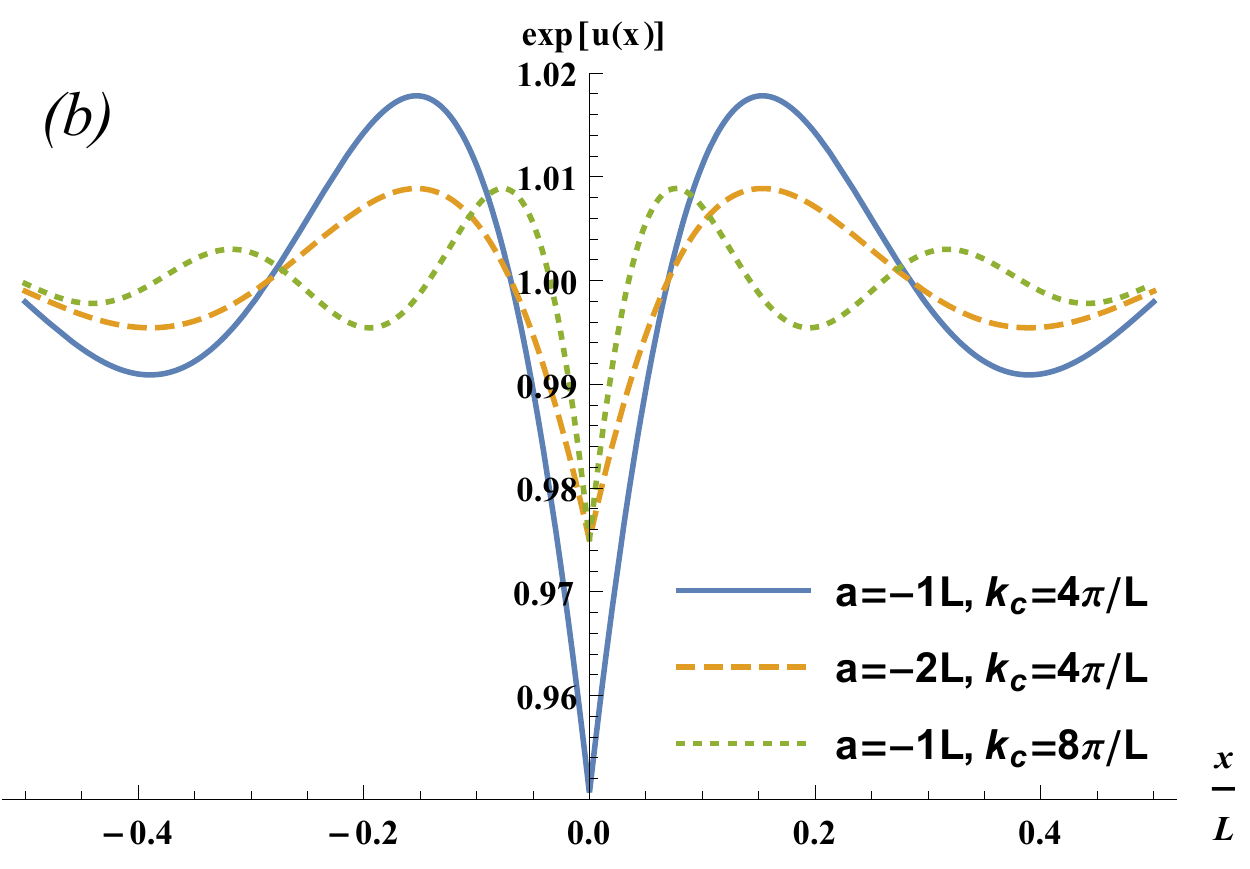}
\caption{Jastrow factor $\mathrm{exp}[u(x)]$ with $u(x)$ from \rref{FTu0OneD}. (a) Attractive and (b) 
repulsive interactions with parameter values for $a_s$ and $k_c$ as indicated. 
\label{Fig:1D}}
\end{figure}

The  Jastrow factor with $u(x)$ from \rref{FTu0OneD} 
is shown in Fig.\ \ref{Fig:1D}.
Close to the coalescence point it resembles the absolute value function, as is expected from \rrefsa{BPOneD} and \rrefsb{JastrowTaylor}. It can be also read from these
equations that the slopes of the two sides of 
the absolute value function linearly depend on the 
inverse of the scattering length. When the scattering length
tends to  infinity the slope goes to zero and the cusp disappears. This is the non-interacting limit.

The parameter $k_c$ adjusts the width of the 
Jastrow factor. As we choose larger momentum cutoff in
the momentum space it makes the function  narrower in  real space. 


The physically relevant information about the scattering length comes exclusively from the cusp of the Jastrow factor near the coalescence point.
The long-range behavior is an artifact from the definition \rrefsb{u1d}. Since the long-range part of the correlation factor is smooth, however,  it is easier to correct it
with the Fock-space expansion of the transcorrelated wave function $\Phi$. Moreover, the long-range part can be easily damped by increasing the parameter $k_c$. In Sec.\ \ref{sec:Numcomp} we will numerically examine the accuracy of the transcorrelated method and we will show that it improves the efficiency of the numerical approach.

\subsection{\label{sec:explicit}Transcorrelated Hamiltonian in real space}

The explicit form of the effective Hamiltonian in \rref{effectiveScheq} can be determined 
from the 
expansion
\begin{eqnarray} \label{eq:bch}
\tilde{H} &=& \hat{H} + \left[ \hat{H}, \tau \right] + 
\frac{1}{2} \left[ \left[ \hat{H}, \tau \right] , \tau \right] + 
\dots \ .
\end{eqnarray}
The external and the particle-particle interaction potentials 
commute with the function $\tau$ as they can be
expressed as a function of the particle positions. The only non-zero 
commutators come from the kinetic term, where only the first three 
terms are different from zero. Thus the expansion \eqref{eq:bch} terminates to yield
\begin{align*}
\tilde{H} = \hat{H} - \frac{\hbar^2}{2m}\sum_{i} \left[ \frac{\partial^2}{\partial x_i^2} , \tau \right] -
\frac{\hbar^2}{4m} \sum_{i} \left[ \left[ \frac{\partial^2}{\partial x_i^2}, \tau \right] , \tau \right] .
\end{align*}
The remaining commutators can be calculated analytically \cite{boys_determination_1969,luo_combining_2018},
\begin{align}
&\tilde{H} = \hat{H} -  \label{effectiveHam} \\
& \hspace{0.2cm} - \sum_{i} \left[ \frac{1}{2} 
\frac{\partial^2 \tau}{\partial x_i^2} +
\frac{\partial \tau}{\partial x_i} 
\frac{\partial}{\partial x_i} +
\frac{1}{2} \left( \frac{\partial \tau}{\partial x_i} \right)^2\right]
\frac{\hbar^2}{m} \ . \nonumber
\end{align}
As it can be seen from
 the term $\frac{\partial \tau}{\partial x_i} 
\frac{\partial}{\partial x_i} $ in the summation on the right-hand side,
the resulting effective Hamiltonian is not Hermitian. As a consequence, the left and right eigenvectors are different and connected by the linear transformation
\begin{eqnarray*}
| \Phi^\mathrm{L} \rangle &=&  e^{2\tau} | \Phi \rangle \ ,
\end{eqnarray*}
where $| \Phi^\mathrm{L} \rangle$ is the adjoint of the left eigenvector and 
$|\Phi \rangle$ is the right eigenvector. 
Since the transcorrelated transformation of Eq.\ \eqref{effectiveScheq} is a similarity transformation, the effective Hamiltonian $\tilde{H}$ nevertheless has the same real-valued eigenvalue spectrum as the original (Hermitian) Hamiltonian $\hat{H}$. 

Properties such as correlation functions or expectation values of general operators can be expressed through the transcorrelated wave function as
\begin{align} \label{eq:expvPhi}
\langle \Psi | \hat{A} |\Psi\rangle &= \frac {\langle\Phi| e^\tau  \hat{A} e^\tau |\Phi\rangle} {\langle\Phi| e^{2\tau}  |\Phi\rangle}.
\end{align}
Evaluating such expectation values involves high-dimensional integrals, which is usually prohibitive in a Fock basis. If the correlation factor is small, however, one can make use of the cumulant expansion
\begin{align} \label{eq:expexpansion}
 \frac {\langle\Phi| e^\tau  \hat{A} e^\tau |\Phi\rangle} {\langle\Phi| e^{2\tau}  |\Phi\rangle} =& \langle\Phi|\hat{A} |\Phi\rangle _c + \langle\Phi|\{\hat{A}, \tau\}|\Phi\rangle _c \nonumber \\& +\frac 1 2 \langle\Phi|\{\{\hat{A}, \tau\},\tau\} |\Phi\rangle _c  + \cdots ,
\end{align}
to obtain approximate results.
Here $\{\cdot ,\cdot\}$ is the anti-commutator and $\langle\Phi|\cdots|\Phi\rangle _c$ denotes the cumulant of operator products \cite{fulde_correlated_2012}.

\subsection{Smoothness of the  transcorrelated wave function \label{sec:smoothness}}

The transcorrelated transformation  improves the smoothness of the wave function, which eventually leads to faster convergence of the basis expansion. Here we consider two interacting particles in a smooth and separable external potential, where we will prove that the transcorrelated eigenfunction is at least $C^1$, i.e.~it can be differentiated at least once with a continuous derivative. In the case of additional even exchange symmetry of the wave function, the smoothness further improves to $C^2$, i.e.~the second derivative of the wave function is also continuous. This is an improvement to the eigenfunctions of the original Hamiltonian which are only $C^0$.

We consider two particles, e.g.~two bosons, or fermions in different spin states, and introduce the center-of-mass and relative coordinates
\begin{align}\label{eq:X}
X & =  \frac{1}{\sqrt{2}}(x_1+x_2) \ ,\\\label{eq:x}
x & =  \frac{1}{\sqrt{2}}(x_1-x_2) \ , 
\end{align}
respectively. For convenience, we assume that the smooth external trapping potential $V(x)$ is further separable, as it is the case for harmonic trapping potentials frequently employed for ultracold atoms:
\begin{align}
\sum_{i=1}^2 V \left( x_i \right) = v(x)+\mathcal{V}(X).
\end{align}
It follows that the Hamiltonian of Eq.~\eqref{scheq} can be written as the sum $H=H_\mathrm{COM} +H_\mathrm{rel}$, with terms that only depend on either the center-of-mass or relative coordinate, respectively. The Schr\"odinger equation \eqref{eq:stationarySE} is thus solved with a wave function of the form
\begin{align} \label{eq:sep}
\Psi(x_1,x_2) = \chi(X) \psi(x) , 
\end{align}
and eigenvalue $E= E_\mathrm{COM} +E_\mathrm{rel}$. The equation for the center-of-mass motion is free of singular operators and thus leads to a smooth wave function $\chi(X)$. The equation for the relative motion, however, contains the particle-particle contact interaction
\begin{align}
\left[ -\frac{\hbar^2}{2m}
\frac{\partial^{2}}{\partial x^{2}}
+ v \left( x \right) + 
\frac{g}{\sqrt{2}} \delta \left( x \right) 
 \right] \psi  =  E_\mathbf{rel}\psi .
\end{align}
The wave function of relative motion $\psi(x)$ is not smooth but rather has a cusp
as we discussed earlier in Sec. \ref{sec:cusp}.

Since we are only interested in the smoothness properties at the particle coalescence point $x=0$, we may take the simplified correlation factor
\begin{align}
\tau(x)=-\frac{\sqrt{2}}{a}|x| \ , \label{relcorrfact}
\end{align}
which has the same small-$x$ expansion up to first order as the function of Eq.~\eqref{FTu0OneD}.
Applying the transcorrelated similarity transformation \rref{effectiveScheq} with $\psi(x) = \exp[\tau(x)] \phi(x)$,
the $\delta$-function interaction term is eliminated.
The transcorrelated Schr\"odinger equation for the relative coordinate then can be written in the form
\begin{align}
-\frac{\hbar^2}{2m}\frac{\mbox{d}^{2} \phi}{\mbox{d}x^{2}}&=\frac{\sqrt{2}\hbar^2}{ma}S(x)\frac{\mbox{d}\phi}{\mbox{d}x}-v(x)\phi+E'\phi \ , \label{eq:1dtc} 
\end{align}
where a constant term from the first derivative of $\tau$ has been absorbed as a shift in the energy $E'=E_{\mathbf{rel}}+{\hbar^2}/{2ma^2}$, and
\begin{align}
S(x)&=\frac{\mbox{d}|x|}{\mbox{d}x}=\begin{cases}
1, & x>0 \ , \\
-1, & x<0 \ . 
\end{cases}
\end{align}

In order to examine the smoothness of the 
function $\phi(x)$, 
we follow the
idea of Kato \cite{kato_eigenfunctions_1957} by
designing an elementary solution of $\frac{\mbox{d}^{2}}{\mbox{d}x^{2}}$,
\begin{align}
G(x)=\frac{|x|}{2}\zeta(x) \ ,
\end{align}
where $\zeta$ is a sufficiently smooth function that equals
1 for $|x|\le1$ and 0 for $|x|>2.$ Then we have
\[
\frac{\mbox{d}^{2}}{\mbox{d}x^{2}}G(x)=\begin{cases}
\delta(x), & |x|<1,\\
S(x)\zeta'(x)+\frac{|x|}{2}\zeta''(x), & \mbox{otherwise}.
\end{cases}
\]
We apply a convolution with respect to $G$ on both sides of
\rref{eq:1dtc}
\[
G*\left(\frac{\mbox{d}^{2}}{\mbox{d}x^{2}}\phi\right)=-2G*\left(\frac{\sqrt{2}}{a}S(x)\frac{\mbox{d}\phi}{\mbox{d}x}-\frac{m}{\hbar^2}v(x)\phi+\frac{mE'}{\hbar^2}\phi\right),
\]
and we find the leading singular term for $\phi$
\begin{eqnarray*}
\phi(x) & = & -\frac{2\sqrt{2}}{a}\int\limits_{-\infty}^{\infty}\frac{|x-y|}{2}\zeta(x-y)S(y)\phi'(y)\mbox{d}y \ + \\ 
&& \hspace{5cm} + \ \mbox{smooth part} \ . 
\end{eqnarray*}
Using partial integration we obtain an integral equation that depends on $\phi(x)$ instead of its derivative
\begin{eqnarray*}
\phi(x) & = & \frac{\sqrt{2}}{a}\int\limits_{-\infty}^{\infty}{\Big [}S(y-x)S(y)\zeta(x-y) \ + \\
&& \hspace{1.5cm} + \ 2|x-y|\delta(y)\zeta(x-y) \ + \\
&& \hspace{1.8cm } + \ |x-y|S(y)\zeta'(x-y){\Big ]}\phi(y)\mbox{d}y \ + \\
&& \hspace{5cm} + \ \mbox{smooth part}.
\end{eqnarray*}
By noticing that $S(y-x)S(y)=1$ in the whole integration domain except $(0,x)$
or $(x,0)$ (depending on whether $x>0$ or $x<0$), we obtain
\begin{align*}
\phi(x) & =\frac{2\sqrt{2}}{a}\begin{cases}
x\zeta(x)\phi(0)-\int\limits_{0}^{x}\phi(y)\zeta(x-y)\mbox{d}y \ + & \\ 
\hspace{3cm} + \ \mbox{smooth part}, & x>0,\\
-x\zeta(x)\phi(0)-\int\limits_{x}^{0}\phi(y)\zeta(x-y)\mbox{d}y \ + & \\
\hspace{3cm} + \ \mbox{smooth part}, & x<0.
\end{cases}
\end{align*}
Hence,  $\phi(x)$ is continuous. Since the singularity only
takes place at $x=0,$ we can simply take $\zeta(x)=1$ for small
variables. Then we get the expressions for the first, second and third 
derivatives ($\phi' \equiv \mathrm{d}\phi/\mathrm{d}x$)
\begin{eqnarray}
\phi'(x)&=&\frac{2\sqrt{2}}{a}\begin{cases}
\phi(0)-\phi(x)+\hspace{1cm}& \\
\hspace{1cm} +\mbox{smooth part}, & x>0,\\
-\phi(0)+\phi(x)+\hspace{1cm}&\\
\hspace{1cm}+\mbox{smooth part}, & x<0,
\end{cases} \label{firstderiv2p} \\
\phi''(x)&=&\frac{2\sqrt{2}}{a}\begin{cases}
-\phi'(x)+\mbox{smooth part}, & x>0,\\
\phi'(x)+\mbox{smooth part}, & x<0,
\end{cases} \label{secondderiv2p}\\
\phi'''(x)&=&\frac{2\sqrt{2}}{a}\begin{cases}
-\phi''(x)+\mbox{smooth part}, & x>0,\\
\phi''(x)+\mbox{smooth part}, & x<0.
\end{cases} \label{thirdderiv2p}
\end{eqnarray}

It follows from Eq.~\eqref{firstderiv2p} that the first derivative $\phi'$ is continuous and thus the relative wave function is $C^1$. Since the center-of-mass wave function $\chi(X)$ is smooth, it follows that also the full wave function $\Psi(x_1,x_2)$ of Eq.~\eqref{eq:sep} is at least $C^1$.

Even stronger results follow when the  wave function  is known to be symmetric under particle exchange, i.e.\ $\Psi(x_1,x_2) =  \Psi(x_2,x_1)$. This is manifestly the case when the particles are spinless bosons but also for the ground state of distinguishable particles if the Hamiltonian is symmetric (e.g.\ as well for~fermions with different spin quantum numbers in the absence of spin-dependent terms in the Hamiltonian). From the symmetry of $\Psi$ it then follows that the relative wave function is even, $\phi(x)=\phi(-x)$, and, as a consequence, its first derivative is an odd function. From \rref{firstderiv2p} it can be seen that the first derivative is also continuous, which means that it must have a node at the origin, i.e.~$\phi'(0)=0$. Using this fact it can be seen from \rref{secondderiv2p} that the second derivative is continuous as well, and the third derivative is the first one where a discontinuity may appear. In this case of even particle exchange symmetry, the transcorrelated wave function is thus $C^2$, i.e.~the smoothness have improved by two orders compared to the original wave function $\Psi(x_1,x_2)$.


These results derived for two particles can be expected to carry over to multi-particle wave functions since the only singular term in the Hamiltonian is two-particle interaction in the form of a Dirac $\delta$. For spinless bosons the multi-particle wave function is symmetric under the exchange of an arbitrary pair of particle coordinates. Hence we expect the transcorrelated multi-particle wave function to be $C^2$ and thus have improved smoothness by two orders compared to the original wave function. We have checked this property by explicitly constructing the two- and three-particle wave functions of the Lieb-Liniger model of  interacting bosons in a one-dimensional box with periodic boundary conditions \cite{lieb_exact_1963}, and found that the transcorrelated wave functions have continuous first and second derivatives while the third derivatives are discontinuous.


For  fermions the wave function has to be antisymmetric under the exchange of fermions, which carries over to an antisymmetry of the spatial wave function under exchange of two-particle coordinates with the same spin (like-spin pairs are thus not affected by the $\delta$ interaction). Pairs of particles with opposite spin are affected by the $\delta$ interaction but,  for more than two (spin-$\frac{1}{2}$) fermions, the wave function is in general not symmetric under the exchange of the coordinates. Thus it is expected that the transcorrelated wave function is $C^1$ with a discontinuous second derivative. We have explicitly constructed the ground-state wave function for three fermions (one spin-$\uparrow$ and two spin-$\downarrow$) in the Yang-Gaudin model of interacting fermions in a box with periodic boundary conditions and verified that the second derivative of the transcorrelated wave function has a discontinuous jump at the coalescence of different-spin particles. We will discuss further numerical evidence for the $C^1$ nature of the transcorrelated three-particle wave function for fermions in Sec.~\ref{sec:threeparticles}.


\subsection{\label{subsec:Convergence rate} Convergence rate for ground-state energy}

In the numerical procedure we have to truncate the many-particle Hilbert space and work with a finite basis. 
Let us consider the case where we truncate the single-particle Hilbert space with a momentum cutoff $\km$ and otherwise perform an exact diagonalization. 
We want to estimate the size of the error $\delta E = \Ex - \Ea$ that is made by replacing the exact energy $\Ex$ by the eigenvalue $\Ea$ obtained in the truncated basis with cutoff $\km$.


Let us write
\begin{align} \label{eq:transSE}
\tilde{H} |\px\rangle = \Ex |\px\rangle ,
\end{align}
for the eigenvalue equation in full Hilbert space and
\begin{align}
\tilde{H}_{PP} | \pa\rangle = \Ea | \pa \rangle
\end{align}
for the approximate,  truncated eigenvalue equation solved by the computer. Here, we have introduced the  truncated Hamiltonian $\tilde{H}_{PP} = P \tilde{H} P$, where $P$ is
the projector onto the $N$-particle linear space spanned by the Fock states constructed from plane waves with momentum $-\km \le k\le \km$. Noting that $\tilde{H}$ is not necessarily Hermitian and has a left eigenvector equation
\begin{align}
\langle \pL| \tilde{H}_{PP} = \Ea \langle \pL | ,
\end{align}
we may obtain an expression for the energy error $\delta E$ from projecting Eq.\ \eqref{eq:transSE} onto $\langle \pL |$.
Simple manipulation yields
\begin{align} \label{eq:deltaE}
\delta E  \langle \pL | \px\rangle = \langle \pL| \tilde{H}_{PQ} |\px\rangle ,
\end{align}
where $\tilde{H}_{PQ} = P \tilde{H} Q$ and $Q = 1-P$ is the projector onto the complement of the projected space, i.e.~where at least one momentum is $|k|>\km$. We may choose $\langle \pL | \px\rangle = 1$ as a normalization condition for the approximate eigenstate and are thus left with evaluating the overlap on the right-hand side of Eq.~\eqref{eq:deltaE}. Let us, for simplicity, consider the situation of Sec.~\ref{sec:smoothness} of two-particles in the relative motion frame and assume that the exact wave function decays with a power-law
\begin{align}
\px(k) \sim k^{-\alpha}
\end{align}
with an integer exponent $\alpha\ge 1$. Then we obtain
\begin{align}
\nonumber
\delta E & = \langle \pL| \tilde{H}_{PQ} |\px\rangle \\ \label{eq:deltaEgen}
& \sim \int_{k_0}^{{\km}} dp \int_{\km}^\infty dq\, \pL(p) q^{-\alpha} \tilde{H}_{pq} ,
\end{align}
where we have replaced the summation of momenta by integrals and the projection operators determine the range of integration. We have applied a small momentum cutoff $k_0$, which is related to the inverse system size. The expression \eqref{eq:deltaEgen} is general enough to apply both to the original exact diagonalisation problem of the Hamiltonian \eqref{scheq} and to the transcorrelated Schr\"odinger equation \eqref{effectiveScheq}.

\subsubsection{Standard method}
The original Hamiltonian \eqref{scheq} for two particles in the relative motion frame
can be written in momentum space with the matrix elements
\begin{align}
H_{pq}&=\frac{\hbar^2 p^2}{m} \delta_{pq}+ g \  ,
\end{align}
where the off-diagonal term results from the short-range interaction  and couples any momenta equally. The diagonal term of the kinetic-energy does not contribute in Eq.~\eqref{eq:deltaEgen} and we are left with
\begin{align}
\nonumber
 &\int_{k_0}^{{\km}} dp \int_{\km}^\infty dq\, \pL(p) q^{-\alpha} \tilde{H}_{pq} \\
 & \quad = 
 \km^{-\alpha+1}\frac{g}{\alpha}\int_{k_0}^{{\km}} dp \,\pL(p) .
\end{align}
Since the $p$ integral is finite, the scaling of the energy error becomes
\begin{align}
\label{eq:scaleorig}
\delta E & \sim\km^{-\alpha+1} \\
& = \km^{-1} , \label{eq:scaleED}
\end{align}
where in the last equality we have used the result from Sec.~\ref{sec:smoothness} that $\alpha=2$ as a consequence of the $C^0$ cusp of the exact wave function with $\delta$ function interactions.

\subsubsection{Transcorrelated method}\label{sec:trcsmooth}

In the transcorrelated approach, the $\delta$ function interaction is removed and replaced by the less singular operator $S(x)\, d/dx$ as discussed in Sec.~\ref{sec:smoothness}. The matrix elements of the transcorrelated Hamiltonian in momentum space become 
\begin{align}
\tilde{H}_{pq}&=\frac{\hbar^2 p^2}{m} \delta_{pq} + 2\sqrt{2}g  \frac{q \sin^2[L(p-q)/4]}{p-q} .
\end{align}
We want to use Eq.~\eqref{eq:deltaEgen} in order to estimate the energy error. For the purpose of the integral the $\sin^2$ factor may be replaced by $\frac{1}{2}$. This yields
\begin{align} \label{eq:errtc}
\delta E & \sim \sqrt{2} g \int_{k_0}^{{\km}} dp \,\pL(p) F(\alpha-1) , 
\end{align}
where the $q$-integral can be separately performed as
\begin{align}
F(n) =  \int_{\km}^\infty dq\, \frac{q^{-n}}{p-q} .
\end{align}
For integer-valued $n\ge 1$ it is easy to show that $F(n)$ has the series representation
\begin{align}
F(n) =  \sum_{\nu = 0}^\infty \frac{p^\nu}{(n+\nu) \km^{n+\nu}}
\end{align}
In order to evaluate the $p$-integral in Eq.~\eqref{eq:errtc} it is relevant to estimate the $p$-dependence of the left eigenfunction $\pL(p)$. Since the left eigenfunction of the transcorrelated Hamiltonian does not benefit from the removal of the cusp by the Jastrow factor it will have the same asymptotics of the original relative wave function, i.e.~$\pL(p)\sim p^{-2}$. Now the integral \eqref{eq:errtc} can be done term by term for the power series. The asymptotic scaling turns out to be dominated by the first term, which gives
\begin{align}
\delta E & \sim\km^{-\alpha+1} \label{energyfromsmoothness}.
\end{align}
This is the same result as the expression \eqref{eq:scaleorig} for the original short-range interaction, i.e.~the scaling of the energy error is completely determined by the large-$k$ asymptotics of the wave function.

Specifically, for the smooth transcorrelated wave function of class ${C}^2$ for the case of completely symmetric wave functions (bosons or fermions with different spin quantum numbers only), we have $\alpha=4$ and thus the expected scaling of the energy error with the momentum cutoff is
\begin{align}
\delta E & \sim\km^{-3} .
\end{align}

\subsection{Transcorrelated Hamiltonian in second quantization and three-body term}\label{sec:secondquant}

In the following we examine the homogeneous system in a discrete plane wave basis. In order to examine the matrix elements of the transcorrelated Hamiltonian, let us rewrite \rref{effectiveHam} in second quantized form \cite{luo_combining_2018},
\begin{eqnarray}
\tilde{H} &=& \frac{\hbar^2}{2m}\sum_{k \sigma }k^2 \, a_{k,\sigma}^{\dagger} \, a_{k,\sigma} \ + \nonumber \\
&& + \ \sum_{\substack{pqk \\ \sigma \sigma'}
} \, T_{pqk} \Theta_{\sigma \sigma'} \,
a_{p-k,\sigma}^{\dagger} \,  a_{q+k,\sigma'}^{\dagger} \, a_{q,\sigma'} \, a_{p,\sigma} \ + \label{secondeffHam} \\
&& + \ \sum_{\substack{pqs \\ kk' \\ \sigma \sigma'}} 
Q_{kk'}   \Theta_{\sigma \sigma'}  a_{p-k,\sigma}^{\dagger} a_{q+k',\sigma}^{\dagger} 
a_{s+k-k',\sigma'}^{\dagger}
a_{s,\sigma'} a_{q, \sigma} a_{p, \sigma} \ , \nonumber
\end{eqnarray}
where $a_{k, \sigma}^\dagger$ create a one-particle plane wave state with
momentum $k$ and spin $\sigma$, $L$ is the length  of the unit cell, and $\Theta_{\sigma \sigma'}=\delta_{\sigma \sigma'}$ for bosons and
$\Theta_{\sigma \sigma'}=1-\delta_{\sigma \sigma'}$ for fermions. The tensors $\mathbf{T}$ and $\mathbf{Q}$ can be expressed explicitly with the correlation factor,
\begin{eqnarray}
T_{pqk} &=& 
\frac{g}{L} + \frac{\hbar^2}{mL} {\Bigg (} k^2 \tilde{u}(k) - (p-q)k\tilde{u}(k)+ \frac{W(k)}{L}  {\Bigg )} \ , \nonumber \\
W(k) &=& \sum_{k'}(k-k')k'\tilde{u}(k-k')\tilde{u}(k') \ , \label{tensorT} \\
Q_{kk'} &=& -\frac{k'k \tilde{u}(k) \tilde{u}(k')\hbar^2}{2mL^2} \ . \nonumber
\end{eqnarray}
The summation in \rref{tensorT} contains infinitely many terms. It can be evaluated exactly. The results and derivations are detailed in the Appendix.

Treating the three-body term in the explicit calculation is cumbersome. In order to improve the numerical efficiency we approximated this term with an effective two-body term. 
For the approximation we considered only the diagonal part of the 
three-body term, where
momentum exchanges are equal to each other ($k=k'$). We can recognize the number operator
($\sum_{s \sigma} a_{s \bar{\sigma}}^\dagger a_{s \bar{\sigma}} $). Its effect can be evaluated in advance,
\begin{align*}
& \sum_{\substack{pqsk\\ \sigma \sigma'}
} Q_{kk}  \Theta_{\sigma \sigma'} a_{p-k,\sigma}^{\dagger} a_{q+k,\sigma}^{\dagger} 
a_{s,\sigma'}^{\dagger}
a_{s,\sigma'} a_{q, \sigma} a_{p, \sigma} | \Phi \rangle = \\
& \hspace{2.5cm} = \sum_{\substack{pqk \\ \sigma }} 
Q_{kk} \mathcal{N}_\sigma a_{p-k,\sigma}^\dagger a_{q+k,\sigma}^\dagger a_{q, \sigma} a_{p, \sigma} | \Phi \rangle \ ,
\end{align*}
where $\mathcal{N}_\sigma=N_\sigma=N-2$ for bosons and $\mathcal{N}_\sigma=N-N_\sigma$ for fermions. This approximation is very closely related to the Random Phase Approximation (RPA) \cite{gaskell_collective_1961,armour_calculation_1980}. 
The approximated Hamiltonian
with only one- and two-body terms can be given in the form:
\begin{eqnarray}
\tilde{H}_{\mathrm{ATB}} &=& \frac{\hbar^2}{2m}\sum_{k \sigma }k^2 \, a_{k,\sigma}^{\dagger} \, a_{k,\sigma} \ + \label{atb}\\
&& + \ \sum_{\substack{pqk \\ \sigma \ne \bar{\sigma}}
} \, T_{pqk} \Theta_{\sigma \sigma'}
a_{p-k,\sigma}^{\dagger} \,  a_{q+k,\bar{\sigma}}^{\dagger} \, a_{q,\bar{\sigma}} \, a_{p,\sigma} \ + \nonumber \\
&& + \ \sum_{\substack{pqk \\ \sigma}}
Q_{kk} \mathcal{N}_\sigma a_{p-k,\sigma}^{\dagger} a_{q+k,\sigma}^{\dagger}  a_{q, \sigma} a_{p, \sigma} \ , \nonumber
\end{eqnarray}
where the "ATB" index in the Hamiltonian stands for "approximate three-body."
In the two-particle case and in the limit 
when the system is noninteracting, this approximation becomes irrelevant, because the three-body term does not have any effects.



\section{\label{sec:Numcomp}Numerical examinations}

\subsection{Methods and implementation}\label{sec:methods}
In this section we study numerically
the homogeneous spin-$\frac{1}{2}$ Fermi gas in one dimension  with Hamiltonian of Eq.\ \eqref{scheq} with $V(x)=0$ 
in a box of length $L$ with periodic boundary conditions (ring configuration)
for two to six particles.
Exact solutions for this system are available using the Bethe ansatz  \cite{yang_exact_1967,gaudin_systeme_1967}, which we use 
to calculate exact reference energies.
We then diagonalize the original Hamiltonian 
\rrefsb{scheq} and the transcorrelated Hamiltonian \rrefsb{atb} with approximated three-body terms.
To this end we use a single-particle basis with $M$ plane waves truncated according to 
\begin{align}
|k|\le \km \equiv\frac{M-1}{2} \frac{2\pi}{L} ,
\end{align}
and construct the full multi-particle Fock basis  with dimension ${M\choose N_\uparrow} {M \choose N_\downarrow}$, where $N_\uparrow$ and $N_\downarrow$ are the spin-up and spin-down particle numbers. 
We then express the Hamiltonian as a matrix in this finite Fock basis and 
numerically obtain the ground-state energy and eigenvector
(often referred to as ``exact diagonalization''). 


We also compare our results with the lattice renormalization approach \cite{castin_simple_2004,ernst_simulating_2011}.
The truncated plane wave basis expansion discretizes real space by creating an underlying reciprocal lattice. 
The lattice renormalization approach then adjusts the potential strength $g$ of the discretized $\delta$ interaction in order to 
recover the correct scattering amplitude for two particles \cite{castin_simple_2004}  or, equivalently, yield the correct two-particle ground-state energy to leading order \cite{ernst_simulating_2011}. In order to apply this approach, one simply replaces the interaction constant $g$ in the Hamiltonian \eqref{scheq} by  the renormalized coupling constant
\begin{eqnarray}
\tilde{g} &=& \frac{g}{{1}+\frac{g}{g_0} \label{tildeg}
} ,
\end{eqnarray}
where 
\begin{align}
g_0 = \frac{M \pi^2\hbar^2}{m L} \approx \frac{\km \pi \hbar^2}{m} .
\end{align}

While the exact diagonalization of the Hamiltonian \eqref{scheq} in the Fock bases with or without renormalized interaction strength can be calculated with any diagonalization algorithm, the 
transcorrelated method has an additional complication due to the non-Hermiticity of the transcorrelated Hamiltonian. We apply power iterations to obtain the ground-state energy and eigenvector
\cite{press_numerical_2007}, 
which can be done for non-Hermitian eigenvalue problems.
The power method can be scaled to very large Hilbert spaces with the stochastic implementation provided by 
the Full Configuration 
Interaction Quantum Monte Carlo \cite{booth_fermion_2009,booth_linear-scaling_2014}. Very recently, this approach was combined  
with the transcorrelated method for the homogeneous electron gas \cite{luo_combining_2018}. 

\subsection{Two particles}\label{sec:two}

\begin{figure}
\includegraphics[scale=0.3]{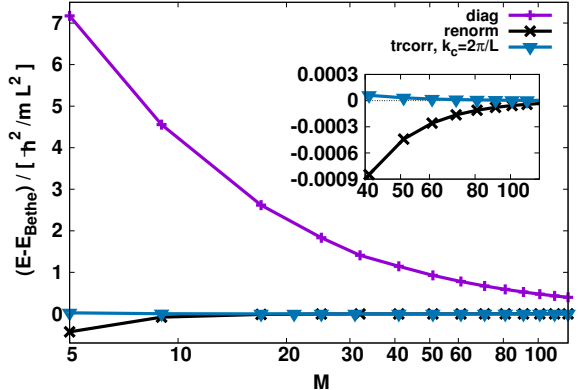}
\caption{{\it Two particles with attractive interaction ($g=-10 \hbar^2/mL$):} the error of the approximate ground-state energy vs.\ the number of one-particle basis functions $M$,
on linear--log scale. ``diag'': exact diagonalization of the Hamiltonian \eqref{scheq}; ``renorm'':  with renormalized interaction constant of Eq.\ \eqref{tildeg}; ``tcorr'': transcorrelated Hamiltonian of Eq.\ \eqref{secondeffHam};  ``$k_{c}$'' truncation parameter for fixed correlation factor of Eq.\ \eqref{u1d}.
\label{fig:absolutetwobodyatt}}
\end{figure}

\begin{figure}
\includegraphics[scale=0.3]{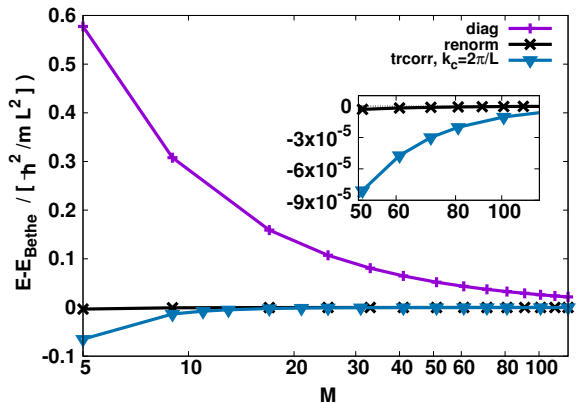}
\caption{{\it Two particles with repulsive interaction ($g=10 \hbar^2/mL$):}
the error of the ground-state energy vs.\ the number of one-particle basis functions $M$ on linear--log scale. Legend labels as in Fig.~\ref{fig:absolutetwobodyatt}.
\label{fig:absolutetwobodyrep}}
\end{figure}

The convergence of the energy with respect to the size of the single-particle basis is shown in  Figs.\ \ref{fig:absolutetwobodyatt} and \ref{fig:absolutetwobodyrep} for two particles (one spin-up and one spin-down fermion or, equivalently, two spinless bosons) with attractive and repulsive interactions, respectively. 
The two-particle system has the advantage that the three-body interaction term in the transcorrelated Hamiltonian of Eq.\ \eqref{secondeffHam} does not contribute and thus the effects of the transcorrelated transformation on the two-particle interactions can be studied in isolation without the need for further approximations.
The energy error $\delta E \equiv E - E_\mathrm{Bethe}$ is calculated as the difference of the numerically obtained approximate value and the exact ground-state energy value obtained from the Bethe ansatz \cite{yang_exact_1967,gaudin_systeme_1967}.
The energy obtained by exact diagonalization of the original Hamiltonian of Eq.\ \eqref{scheq} in the truncated Fock basis (``diag'') is found to converge linearly with the inverse number $M^{-1}$ of one-particle basis functions in Figs.~\ref{fig:twobodyattfixkc} and \ref{fig:twobodyrepfixkc}, which agrees with the theoretical prediction of Eq.\ \eqref{eq:scaleED} in Sec.\
\ref{subsec:Convergence rate}. 

The transcorrelated approach (``trcorr'') is seen to generally improve upon the exact diagonalization results. From Sec.\ \ref{subsec:Convergence rate} we also may expect a faster convergence rate of $\delta E \sim M^{-3}$ (since the two-particle ground-state wave function is symmetric under particle exchange).
From the numerical results presented in Figs.\ \ref{fig:twobodyattslidingkc} and \ref{fig:twobodyrepslidingkc} we see that this is the case asymptotically for basis sets that are large enough to resolve the modified singular feature of the transcorrelated wave function.

\subsubsection{Correlation factor with fixed parameter $k_c$}

\begin{figure}
\includegraphics[scale=0.3]{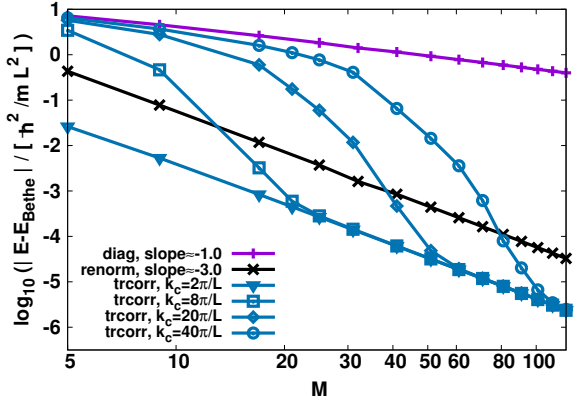}
\caption{{\it Power-law convergence with fixed correlation factor - two particles with attractive interaction ($g=-10 \hbar^2/mL$):} 
the error of the ground-state energy vs. the number of single-particle basis functions M on a log-log scale for different values of the $k_c$ parameter (“trcorr”). Exact diagonalization (“diag”) and renormalized results (“renorm”) are shown for comparison (cf. Fig.~\ref{fig:absolutetwobodyatt}). “slope”: approximate exponent of fitted power-law 
$\delta E\sim M^{\mbox{slope}} $ . 
\label{fig:twobodyattfixkc}}
\end{figure}

\begin{figure}
\includegraphics[scale=0.3]{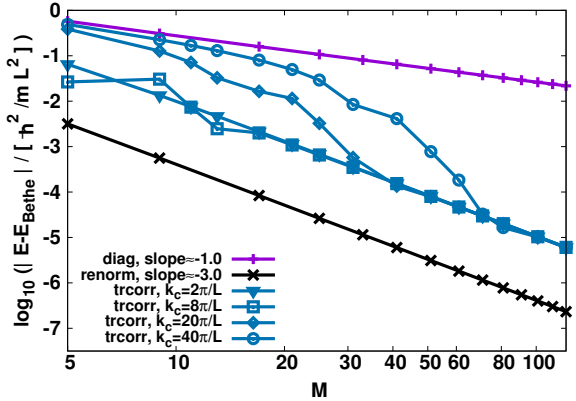}
\caption{{\it Power-law convergence with fixed correlation factor - two particles with repulsive interaction ($g=10 \hbar^2/mL$):}
the error of the ground-state energy vs.\ the number of one-particle basis functions $M$ in a log--log plot. Legend labels as in Fig.~\ref{fig:twobodyattfixkc}.
\label{fig:twobodyrepfixkc}}
\end{figure}

In Figs.~\ref{fig:twobodyattfixkc} and \ref{fig:twobodyrepfixkc} the data labeled with $k_{c}$  values are obtained with fixed correlation factors and variable number of single-particle basis functions $M$.
The smallest value, $k_{c}= 2\pi/L$, shows significantly improved energy errors following the power law $\delta E \sim M^{-3}$ for all considered basis set sizes $M\ge 5$. Increasing the correlation factor cutoff $k_{c}$ leads to an overall smaller correlation factor due to fewer Fourier components contributing, and an associated smaller length scale  for its real-space version $u(x)$ of Eq.\ \eqref{FTu0OneD}. Unsurprisingly, the smaller correlation factors are less effective in reducing the energy error for the small (fixed-size) basis sets. However, when the number of single-particle functions $M$ is increased, all curves collapse onto the same asymptotic power-law with $\delta E \sim M^{-3}$. From our numerical data  we find that the correlation factor is fully effective when $\km  \gtrapprox 3 k_{c}$ for the attractive case of Fig.\ \ref{fig:twobodyattfixkc} and  $\km  \gtrapprox 2 k_{c}$ for repulsive interactions as seen in Fig.\  \ref{fig:twobodyrepfixkc}, where $\km  = (M/2-1) 2\pi/L$ determines the number of single-particle functions $M$.

\subsubsection{Correlation factor with sliding parameter $k_c$}

\begin{figure}
\includegraphics[scale=0.3]{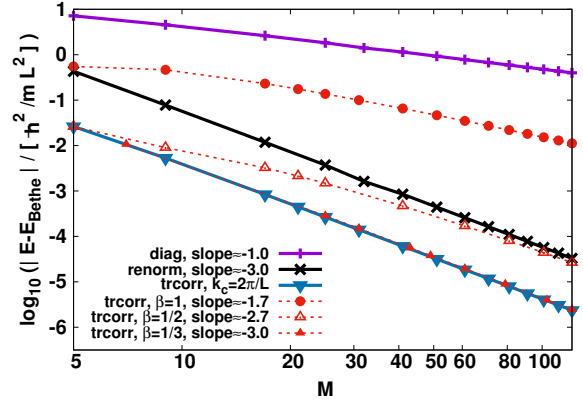}
\caption{{\it Correlation factor with sliding parameter $k_c$ - two particles with attractive interaction ($g=-10 \hbar^2/mL$) with sliding $k_c$:} the error of the approximate ground-state energy vs.\ the number of one-particle basis functions $M$ in a log--log plot. Legend labels as in Fig.~\ref{fig:twobodyattfixkc}. ``$\beta$'' parameter for scaled correlation factor of Eq.\ \eqref{betadef}.  
\label{fig:twobodyattslidingkc}}
\end{figure}

\begin{figure}
\includegraphics[scale=0.3]{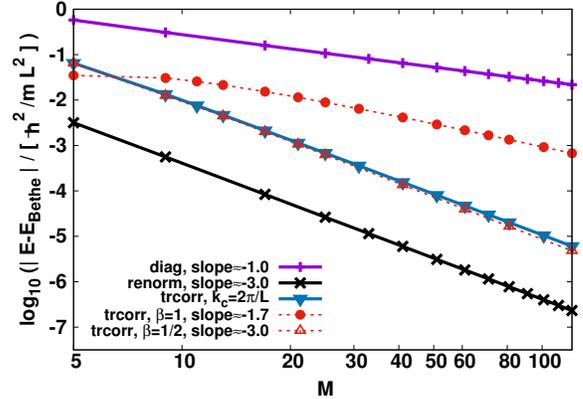}
\caption{{\it Correlation factor with sliding parameter $k_c$ - two particles with repulsive interaction ($g=10 \hbar^2/mL$) with sliding $k_c$:}
the error of the ground-state energy vs.\ the number of one-particle basis functions $M$ in a log--log plot. Legend labels as in Figs.~\ref{fig:twobodyattfixkc}. ``$\beta$'' parameter for scaled correlation factor of Eq.\ \eqref{betadef}. 
\label{fig:twobodyrepslidingkc}}
\end{figure}

The observation that the correlation factor is fully effective when $k_\mathrm{max}$ is larger than a value determined by $k_c$ suggests that it makes sense to adjust $k_c$ with the size of the basis set $M$ (or, equivalently, $\km$), in order to find the smallest correlation factor necessary, for given basis set size $M$, to fully reap the benefits of the accelerated convergence of the transcorrelated approach.
We thus introduce a way of scaling the correlation factor cutoff  $k_{c}$ with the basis set size according to 
\begin{align}
k_c = \beta \km  \ \label{betadef} .
\end{align}
A similar way of scaling the correlation factor with the size of the single-particle basis set was previously  suggested in Ref.\ \cite{luo_combining_2018} (with fixed $\beta={1}$) in order to control the size of the three-body term in the transcorrelated Hamiltonian. We will discuss this issue in Secs.\ \ref{sec:threeparticles} and \ref{sec:sixparticles}. In Figs.\ \ref{fig:twobodyattslidingkc} and \ref{fig:twobodyrepslidingkc} we show, respectively, the energy error obtained with this approach for different values of $\beta$. Our data suggest that 
each value of $\beta$ leads to a different power-law, until the value of $\beta$ is small enough to reach the theoretical limit with $\delta E \sim M^{-3}$. Reducing the value of $\beta$ further, does not change the power-law. We also see that different values of $\beta$ are needed to reach the theoretical limit depending on the nature of the interaction. 

\subsubsection{Comparison with the renormalization approach}

Energies obtained with renormalized interaction strength according to Eq.\ \eqref{tildeg}
are also shown in Figs.\ \ref{fig:absolutetwobodyatt} - \ref{fig:twobodyrepslidingkc} for comparison. The renormalization method works well for estimating the ground-state energy for two particles, which is not surprising because the renormalized coupling constant is determined by comparison with an exact solution to a two-particle problem. We here find that the energy error scales again as $\delta E \sim M^{-3}$. Even though the scaling is the same as the transcorrelated method, we find that the prefactors are different (leading to different off-sets of the curves in Figs.\ \ref{fig:twobodyattslidingkc} and \ref{fig:twobodyrepslidingkc}). It is interesting to note that the transcorrelated approach works better than the renormalized one for attractive interactions but worse for repulsive interactions. This is probably due to the Jastrow factor resembling the bound-state wave function that dominates the ground state for attractive interactions (see Fig.\ \ref{Fig:1D}). Since the prefactor of the transcorrelated energy error certainly depends on the details of the cutoff procedure used in Eq.\ \eqref{u1d}, it could probably be further reduced by optimizing this procedure.


\subsubsection{Single-particle momentum density}

\begin{figure}
\includegraphics[scale=0.3]{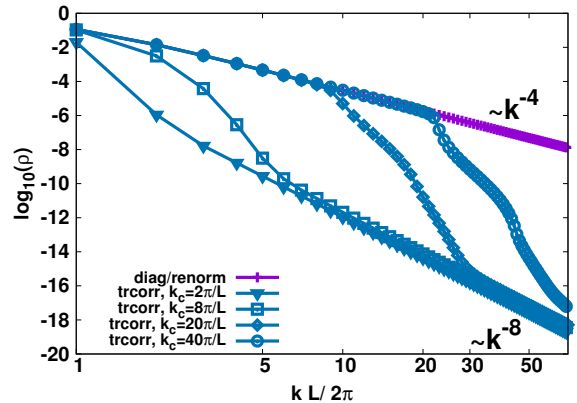}
\caption{Single particle momentum density  $\rho(k)$ for two particles with attractive interactions $g=-10 \hbar^2/mL$, with $M=139$ on a log--log scale showing the transition from the $k^{-4}$ behavior of the original Hamiltonian to the $k^{-8}$ asymptotics of the effective Hamiltonian as an effect of the correlation factor. 
Legend labels as in Fig.~\ref{fig:twobodyattfixkc}. Asymptotic power-laws from approximate fits to the large-$k$ tails as indicated. The data from exact diagonalization with the bare interaction (``diag'') and renormalized interaction (``renorm'') are indistinguishable on the scale of the plot.
\label{fig:twobodydens}}
\end{figure}

In order to obtain 
information about the approximate
wave function, we calculate the 
single-particle momentum density
\begin{align}
\rho_\sigma(k) = \langle a^\dag_{k,\sigma} a_{k,\sigma} \rangle , 
\end{align}
where the expectation value is taken with respect to the finite basis set approximation of either the original ground-state $|\Psi\rangle$ or the transcorrelated eigenstate $|\Phi\rangle$, respectively. 

The results for two particles, shown in Fig.\ \ref{fig:twobodydens}, are independent of the spin $\sigma$ and the sign of $k$. The momentum distribution of the original ground-state  $|\Psi\rangle$ shows a clear power-law decay $\sim k^{-4}$ for almost the entire momentum interval shown in the figure. As discussed in Sec.\ \ref{sec:cusp}, this behavior is expected, since the momentum density contains the square of the wave function, which possesses $k^{-2}$ tails as a consequence of the cusp in real space \cite{olshanii_short-distance_2003}. 
The renormalization method leads to the same power-law for the momentum density, since the analytic properties of the wave function are not changed.

The momentum distributions of the transcorrelated ground-state $|\Phi\rangle$ are seen to decay much faster for large $k$ and asymptotically  converge to a power-law of $k^{-8}$.
This observation is consistent with  the analytic results about the smoothness of the transcorrelated  wave function of Sec.\ \ref{sec:smoothness}. 
Improving the smoothness (differentiability class) of the real-space wave function by one order also decreases the power-law of the large-$k$ tail in momentum space by one order. Thus the $C^2$ character of the transcorrelated two-particle wave function implies $\sim k^{-4}$ scaling of the wave function and thus $\sim k^{-8}$ scaling of the momentum density. The parameter $k_c$ defines an inverse length scale characterizing the ``size'' of the correlation factor. For larger length scales the correlation factor does not have any significant effect and hence the momentum density follows the original $k^{-4}$ power-law for $k<k_c$. For smaller length scales (larger $k$) there is a transition region after which the smoothing effect of the correlation factor on the transcorrelated wave function becomes fully effective. In this regime of the smallest length scales (large $k$), the short-range correlations are suppressed and the momentum density shows a  $~1/k^8$ decay.
Smaller correlation factors (corresponding to larger $k_c$) reach the asymptotic scaling at larger wave numbers, which is expected because the wave-length of the basis functions needs to be small enough to resolve the features of the smaller correlation factor in order to take advantage of the improved smoothness of the wave function.

A remarkable feature of Fig.\ \ref{fig:twobodydens} is that the momentum density of the transcorrelated wave function for $k_c = 20 \pi/L$ and  $k_c = 40 \pi/L$ coincides with the momentum density of the original wave function accurately for the smaller values of $k$ up to critical value that is approximately given by $k_c$. This means that the exact momentum density can be extracted from  Eq.\ \eqref{eq:expvPhi} for the small wave numbers already from the first term of the expansion \eqref{eq:expexpansion}, i.e.
\begin{align}
\langle \Psi | a^\dag_k a_k |\Psi\rangle \approx \langle\Phi| a^\dag_k a_k |\Phi\rangle .
\end{align}
Increasing $k_c$ will further increase the range of wave numbers (equivalently decrease the length scale) over which the momentum density is accurately approximated.

\subsection{Three fermions \label{sec:threeparticles}}

In order to study the role of the  three-body term  and the effects of approximations we need to consider more than two particles. It is also interesting to study the efficacy of renormalizing the interaction strength in a multi-particle system, as Eq.\ \rrefsb{tildeg} was derived considering only two interacting particles.


\begin{figure}
\includegraphics[scale=0.3]{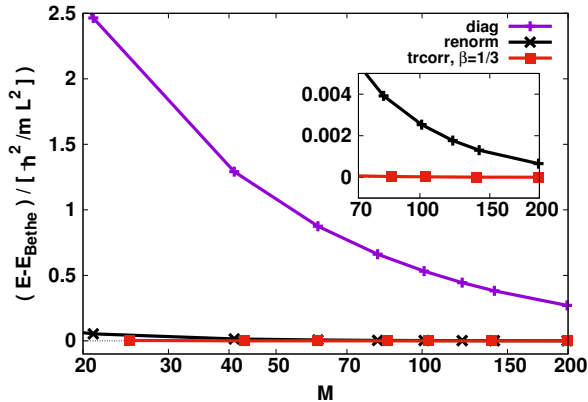}
\caption{{\it Three fermions (two spin-up and one spin-down) with attractive interactions $g=-10 \hbar^2/mL$:} the  error of the ground-state energy vs.\ the number of one-particle basis functions $M$ on a linear-log scale. The labels ``diag'', ``renorm'' and ``tcorr'' stand for the exact diagonalization, renormalization, and transcorrelated approaches, respectively. The correlation factor cutoff $k_c$ is  linearly scaled with $M$ according to Eq.\ \eqref{betadef}.  
\label{threebodyatt}}
\end{figure}

We consider three spin-$\frac{1}{2}$ fermions with two spin-up and one spin-down particles with attractive interactions.
The energy error compared to the exact Bethe-ansatz solution for the various approximations is shown in Fig.\ \ref{threebodyatt}. Here (and in other figures),  the lines connecting data points are a guide to the eye only. Where connecting lines are omitted in the logarithmic plot, a sign change of the error has occurred, i.e.\ the approximate energy curve crosses the exact one.
The energy error from the exact diagonalization of the original Hamiltonian \eqref{scheq} is shown for reference in Figs.\ \ref{threebodyatt} and \ref{threebodyattslidingkc}a. As expected it follows the power-law scaling $\sim k^{-1}$. The approach of renormalizing the interaction strength for a given basis set size by Eq.\ \eqref{tildeg} is shown in Figs.\ \ref{threebodyatt} and \ref{threebodyattslidingkc}a, and clearly demonstrates power-law scaling $\sim M^{-2}$. The convergence rate has decreased by one order compared to the two-particle case. This can be understood by the fact that the renormalized interaction strength was determined by solving a two-particle problem. 

\subsubsection{Correlation factor with fixed parameter $k_c$: Bias from the approximation of the three-body term}

\begin{figure}
\includegraphics[scale=0.3]{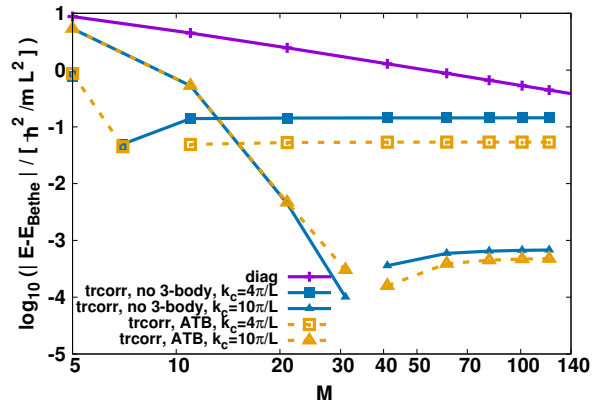}
\caption{{\it Relevance of three-body terms - three fermions (two spin-up and one spin-down) with attractive interactions $g=-10 \hbar^2/mL$ with constant $k_c$:} the  error of the ground-state energy vs.\ the number of one-particle basis functions $M$ in a log-log plot. Legend labels as in Fig.~\ref{threebodyatt}. The label ``no 3-body'' means that the three-body term was omitted from the transcorrelated Hamiltonian \rref{secondeffHam}, while for ``ATB''-labelled data the three-body term is approximated as per \rref{atb}. 
\label{threebodyattfixedkc}}
\end{figure}

Results from the transcorrelated approach with fixed cutoff parameter $k_c$ are shown in Fig.\ \ref{threebodyattfixedkc}. Since we are not including the full three-body terms in our diagonalization procedure, the results converge to a finite value, which quantifies the contribution of the neglected three body terms. It can be seen that the approximate inclusion of the three-body term as per Eq.\ \eqref{atb} (labelled ``ATB'') leads to smaller errors than the complete neglect of three-body contributions [``no 3-body''; Eqs.\ \eqref{secondeffHam} and \eqref{tensorT} with $Q_{kk'}=0$].
By increasing
the value of $k_c$, the three-body error decreases dramatically providing a more accurate approximation for the energy. This can be understood as follows: increasing  $k_c$ reduces the length scale associated to the correlation factor and with it the range of the newly generated terms in the effective Hamiltonian, including the three-body term. In a dilute gas, the significance of the three-body terms thus diminishes.

\subsubsection{Correlation factor with sliding parameter $k_c$: Treatment of the three-body term}

\begin{figure}
\includegraphics[scale=0.3]{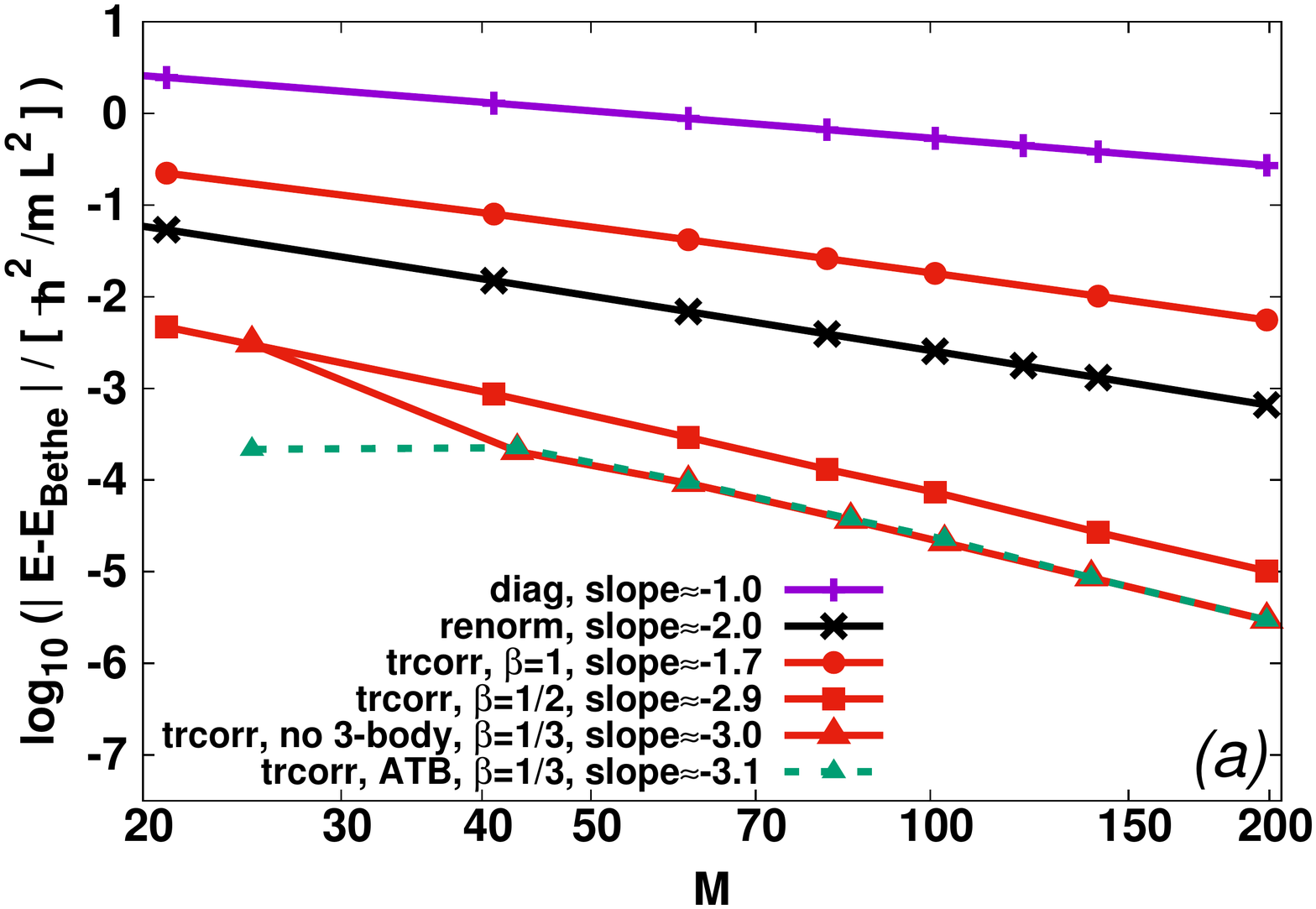}
\includegraphics[scale=0.3]{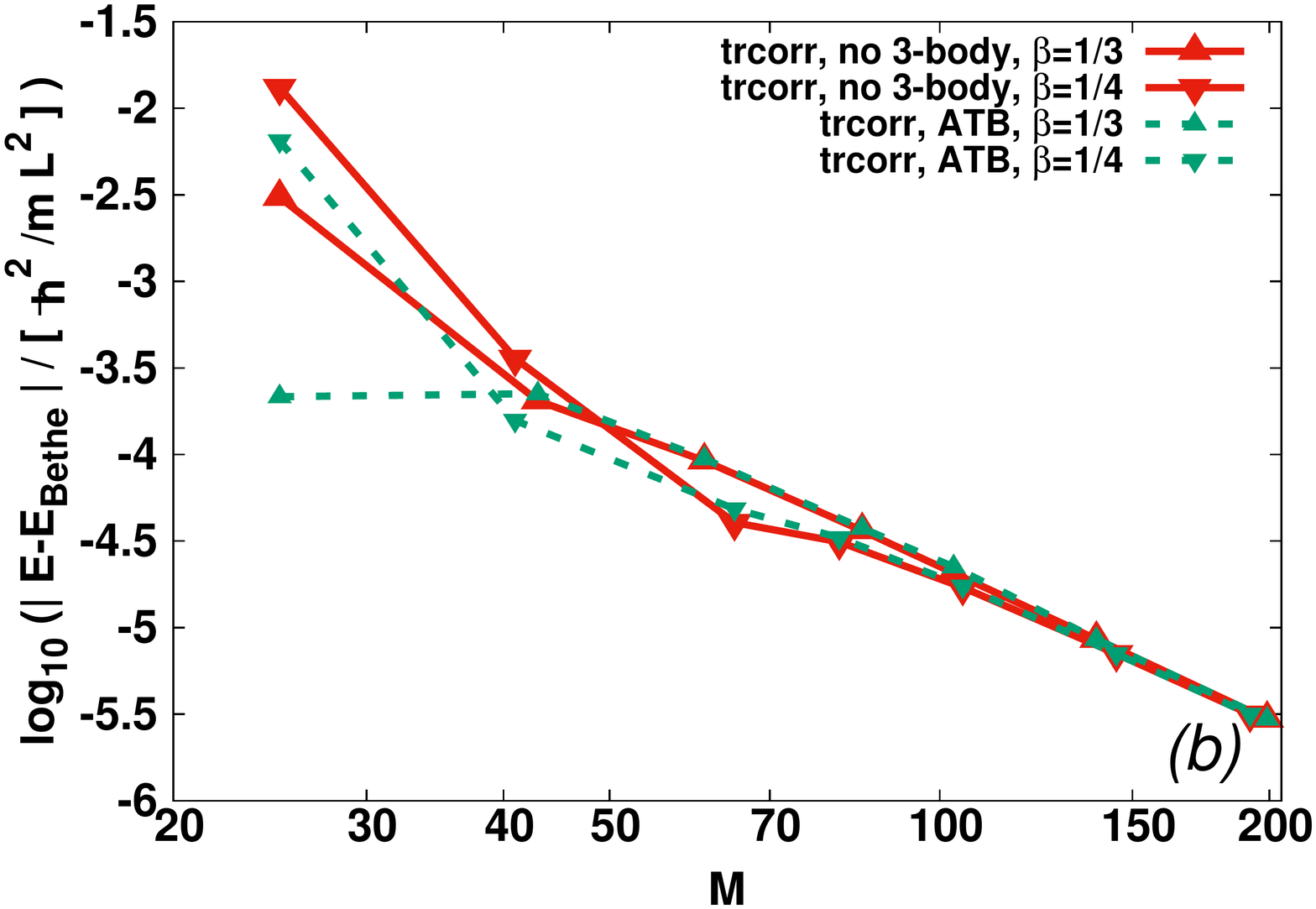}
\caption{{\it Power-law convergence with sliding correlation parameter $k_c$ - three fermions (two spin-up and one spin-down) with attractive interactions $g=-10 \hbar^2/mL$:} the  error of the ground-state energy vs.\ the number of one-particle basis functions $M$ in a log-log plot. Legend labels as in Fig. ~\ref{threebodyatt}. The label ``no 3-body'' means that the three-body term was omitted from the transcorrelated Hamiltonian \rref{secondeffHam}, while for ``ATB''-labelled data the three-body term is approximated as per \rref{atb}. Where these labels are not indicated  the results of the two approaches are indistinguishable from each other on the scale of this plot.   
\label{threebodyattslidingkc}}
\end{figure}

We may  expect that scaling the parameter $k_c$ of the correlation factor with the size of the basis set as per Eq.\ \eqref{betadef} is a way to asymptotically eliminate the error introduced by neglecting or approximating the three-body term and converge to exact results. Figs.\ \ref{threebodyatt} and \ref{threebodyattslidingkc} show that this is indeed the case (and a similar observation was previously made in Ref.\ \cite{luo_combining_2018}). Data for different values of the scaling factor $\beta$ in Eq.\ \eqref{betadef} all show algebraic convergence to the exact ground-state energy. The numerically extracted power-law exponents vary, with generally a smaller value of $\beta$ resulting in faster convergence in the asymptotic (large $M$) regime. As in the case for two particles in Fig.\ \ref{fig:twobodyattslidingkc}, the fastest convergence is reached with $\beta \le \frac{1}{3}$ yielding the approximate power law $\delta E\sim M^{-3}$. As seen in Fig.~\ref{threebodyattslidingkc}b, decreasing the factor $\beta$ below this value does not yield a further improvement of the asymptotic power law, but on the other hand, leads to larger errors for smaller basis sets (due to the smaller correlation factor being less effective in capturing pair correlations). It is also seen from the data in Fig.~\ref{threebodyattslidingkc}  that the approximate treatment of the three-body term  (``ATB'') of Eq.\ \eqref{atb}  does not change the asymptotic power law, or even the value of the energy error in the asymptotic regime, but it does improve the energy error for smaller basis sets. We conclude that the value of $\beta = \frac{1}{3}$ and the inclusion of approximate three-body terms gives the best performance.

\subsubsection{Single-particle momentum density}

\begin{figure}
\includegraphics[scale=0.3]{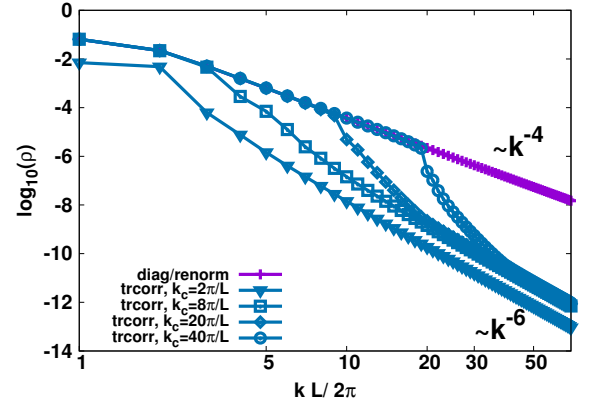}
\caption{ Single particle momentum density  $\rho_\downarrow(k)$ for  the minority spin component for three fermions (two spin-up, one spin-down) with attractive interactions of $g=-10 \hbar^2/mL$ and $M=139$ on a log-log scale.  
Labels as in Fig.~\ref{threebodyatt}.  Asymptotic power laws from approximate fits to the large-$k$ tails as indicated. The data from exact diagonalization with the bare interaction (``diag'') and renormalized interaction (``renorm'') are indistinguishable on the scale of the plot.
\label{fig:threebodydens}}
\end{figure}

The single-particle momentum density $\rho_\downarrow(k)$  is shown in Fig.\ \ref{fig:threebodydens} and shows similar features as seen in the two-particle case of Fig.\ \ref{fig:twobodydens}.
The original ground-state wave function for three fermions leads to a $\sim k^{-4}$ algebraic decay of the momentum density as in the case of two particles, or more generally, for the Bose gas \cite{olshanii_short-distance_2003}. 
The transcorrelated  ground-state $|\Phi\rangle$, however, asymptotically decays as  $\sim k^{-6}$, which is  slower by two orders than in the two particle case. This observation suggests that the wave function has the differentiability class $C^1$,  i.e.\ is less smooth by one order than the two-particle wave function. This result provides further evidence for the conclusion of Sec.\ \ref{sec:smoothness}  that the transcorrelated fermionic multi-particle wave function is $C^1$.

The remarkable result from the numerical investigation of the three-fermion system is that the ground-state energy convergence $\sim M^{-3}$ is faster than expected from the analytical estimates of Sec.\ \ref{sec:trcsmooth}. The analytical arguments as well as the numerical analysis of the momentum density of the transcorrelated wave function indicate that the large momentum asymptotics scale as $\Phi(k)\sim k^{-3}$, which, by Eq.\ \eqref{energyfromsmoothness}, should lead to an energy error scaling as $\delta E  \sim M^{-2}$. The faster-than-expected convergence of the transcorrelated energy is well supported by the data shown in Fig.\ \ref{threebodyattslidingkc} and comes as a pleasant surprise.


\subsection{Six fermions} \label{sec:sixparticles}

\begin{figure}
\includegraphics[scale=0.3]{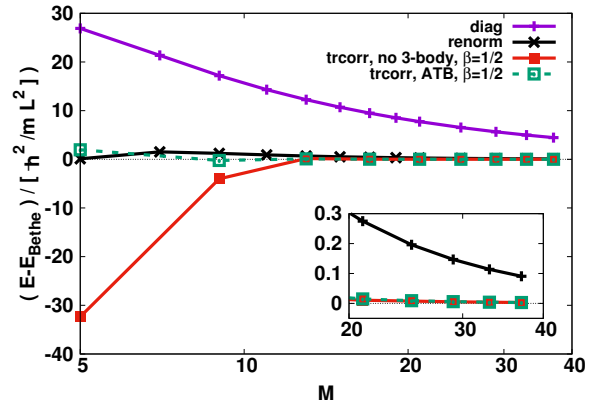}
\caption{{\it Six fermions (three spin-up and three spin-down) with attractive interactions $g=-10 \hbar^2/mL$:} the  error of the ground-state energy vs.\ the number of one-particle basis functions $M$ on a linear-log scale. Labels as in Figs.\ \ref {threebodyatt} and \ref{threebodyattslidingkc}.
\label{sixbodyatt}}
\end{figure}

\begin{figure}
\includegraphics[scale=0.3]{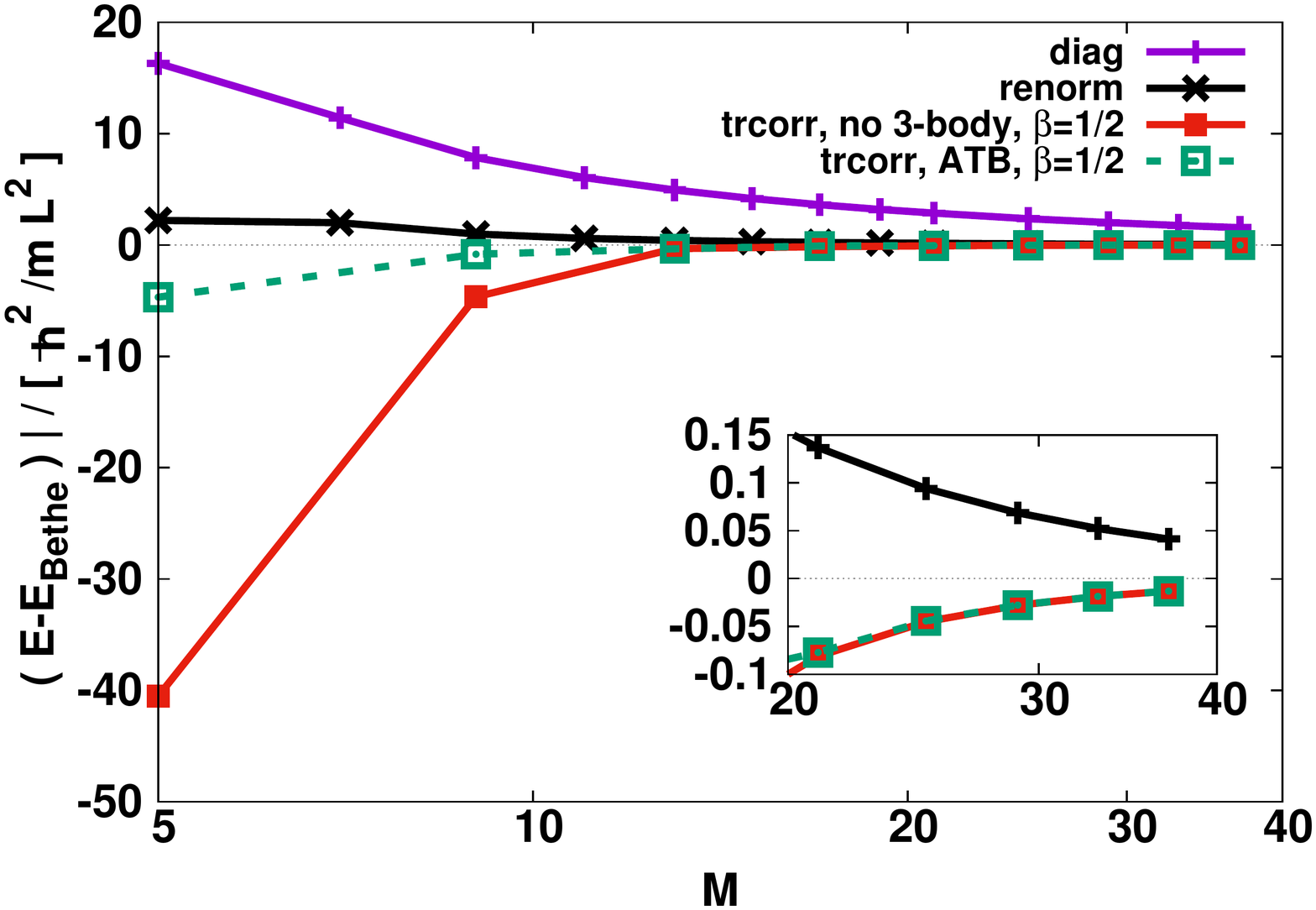}
\caption{{\it Six fermions (three spin-up and three spin-down) with repulsive interaction $g=10 \hbar^2/mL$:}  the  error of the ground-state energy vs.\ the number of one-particle basis functions $M$ on a linear-log scale. Labels as in Figs.\ \ref {threebodyatt} and \ref{threebodyattslidingkc}.
\label{sixbodyrep}}
\end{figure}

\begin{figure}
\includegraphics[scale=0.3]{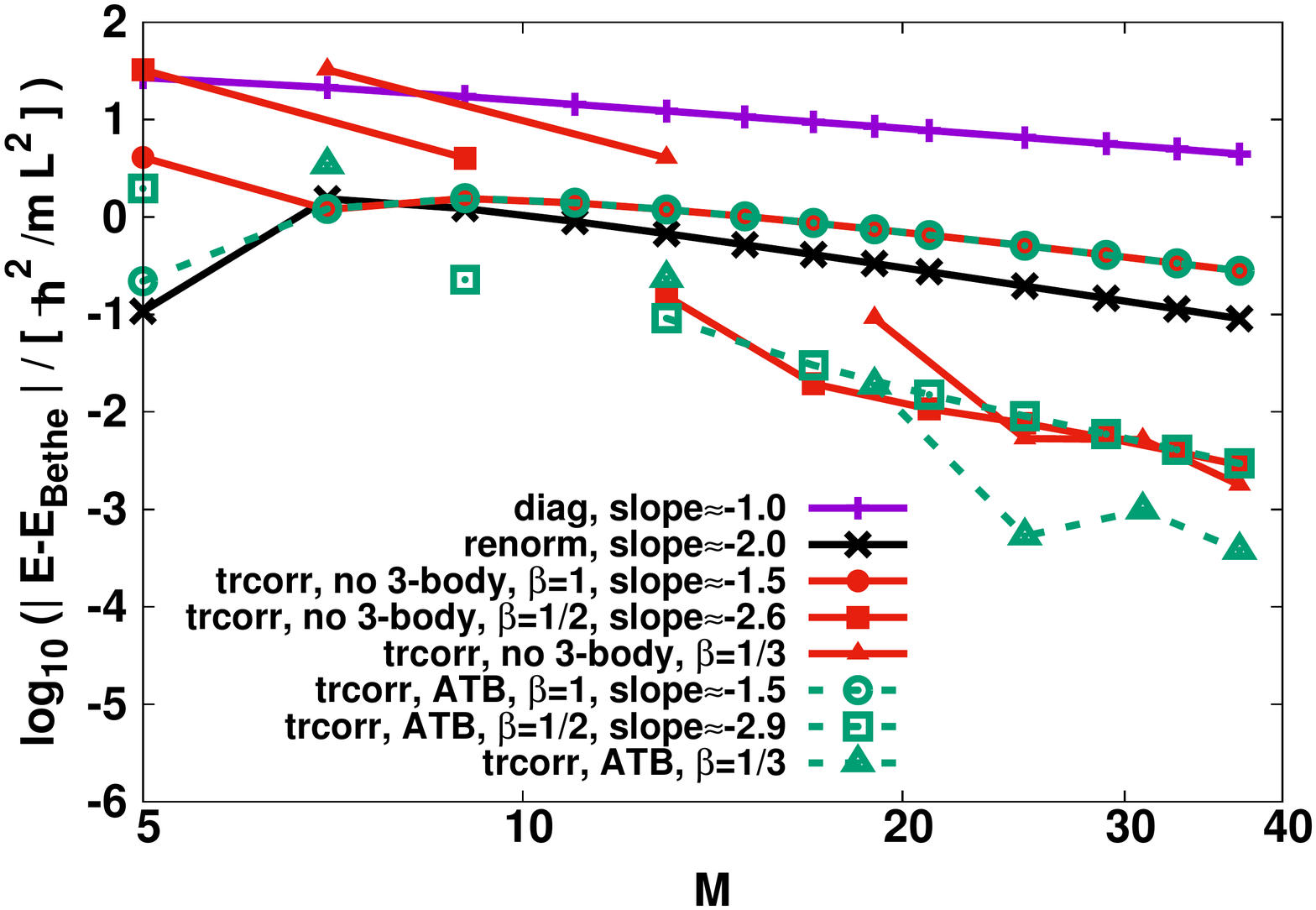}
\caption{{\it Power-law convergence with sliding correlation parameter $k_c$ - six fermions (three spin-up and three spin-down) with attractive interactions $g=-10 \hbar^2/mL$:} the  error of the ground-state energy vs.\ the number of one-particle basis functions $M$ on a log-log scale. Labels as in Figs.\ \ref {threebodyatt} and \ref{threebodyattslidingkc}.
\label{sixbodyattscaling}}
\end{figure}

\begin{figure}
\includegraphics[scale=0.3]{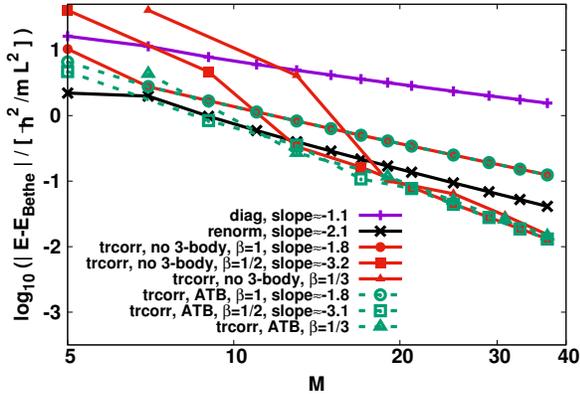}
\caption{{\it Power-law convergence with sliding correlation parameter $k_c$ - six fermions (three spin-up and three spin-down) with repulsive interaction $g=10 \hbar^2/mL$:}  the  error of the ground-state energy vs.\ the number of one-particle basis functions $M$ on a log-log scale. Labels as in Figs.\ \ref {threebodyatt} and \ref{threebodyattslidingkc}.
\label{sixbodyrepscaling}}
\end{figure}

We also examined the spin-balanced six-particle system  in order to study the convergence properties for the larger particle number. Figs.\ \ref{sixbodyatt}-\ref{sixbodyrepscaling} show the convergence of the energy error for attractive and repulsive interactions, respectively. The picture for the convergence rates of the different approximations is largely consistent with the results for three fermions. Exact diagonalization of the original Hamiltonian yields an $M^{-1}$ convergence, as expected, and also the convergence rate of $M^{-2}$ for the renormalization approach has not changed compared to three particles. This confirms that the faster convergence of the renormalization method in the two-particle system is a special case.

The transcorrelated approach with scaled correlation factor cutoff $k_c$ is also here seen to converge algebraically towards the exact ground-state energy, even though the three-body terms have been either approximated or fully omitted. The results indicate that a scaling factor of $\beta = \frac{1}{2}$ is sufficient to reach optimal convergence of the energy error. This indicates that the required $\beta$-factor for optimal convergence has no strong dependence on the particle number. The six-particle results also confirm the faster-than-expected $M^{-3}$ scaling of the transcorrelated approach that we already saw for the three-particle and two-particle cases. 


\section{\label{sec:Conclusion}Conclusion and outlook}
We have constructed an effective Hamiltonian based on a transcorrelated transformation that replaces the singular contact interaction by less singular, but non-Hermitian terms, which also include artificial three-body interactions. 
We have shown that an explicit treatment of the three-body terms can be avoided, while nevertheless achieving fast convergence to the exact results by scaling the correlation factor with the size of the single-particle basis. This scaling is controlled with the parameter $\beta$ through \rref{betadef} and effectively reduces the error due to neglected three-body terms when the basis set is increased by reducing the size of the correlation factor. While small values of $\beta$ will lead to the optimal asymptotic scaling of the error for large basis sets, larger values will lead to improved benefits of the correlation factor for smaller basis set but compromise the asymptotic scaling beyond an optimal value. The optimal value of $\beta$ was seen to depend weakly on the sign of the interaction strength and may also depend on the value.

Including the full three-body terms would allow one to achieve unbiased results for a Fock-space diagonalization even for fixed-size correlation factors. This could potentially be useful for reducing the amount of correlation in the effective Hamiltonian, which may be beneficial for the convergence of stochastic or approximate approaches.
Our numerical results indicate that the transcorrelated method
improves the convergence of the energy from $M^{-1}$ of the exact diagonalization of the original Hamiltonian to $M^{-3}$, where $M$ is the number of single-particle basis functions. For two particles we could show that the faster convergence rate originates in the  improved smoothness of the transcorrelated wave function from $C^0$ of the original cusp-like wave function to $C^2$. For more than two spin-$\frac{1}{2}$ fermions, where the smoothness of the wave function improves only to $C^1$, the convergence rate is not yet fully understood.

We have also examined an alternative approach based on a simple renormalization of the interaction constant. While not affecting the smoothness of the wave function, this approach improves the convergence rate of ground-state energies by one order to $M^{-2}$.

Based on such promising results for the ground state of the one-dimensional homogeneous gas, it will be interesting to examine the efficiency of the transcorrelated approach in a trapped system or for excited states, where the presence of a cusp at the two-particle coalescence causes  slow convergence  \cite{grining_many_2015}.  Due to the non-Hermitian nature of the transcorrelated Hamiltonian, care must be taken when choosing an appropriate excited state method, but exact diagonalization approaches, for example, are well suited.
In future work we also would like to investigate two- and three-dimensional systems, 
 as well as extending the treatment of
the three-body terms to include all six-index interaction terms, where we 
expect improvements even more significant than those found
in the present study.

\section{Acknowledgement}
We wish to acknowledge  Elke Pahl and Ulrich Ebling for discussions  and Sophie Shamailov for providing us with Bethe ansatz
energies. The work was supported by the Marsden fund of New Zealand (MAU1604).

\appendix





\section{\label{app:infinitesum}Evaluation of the infinite sum in
\rref{tensorT}}
The infinite summation is easier to evaluate if we make the substitutions
 $k=2\pi n/L$, $k'=2\pi n'/L$ and
$k_c=2\pi n_c/L$ into the sum as
\begin{eqnarray}
W\left(\frac{2\pi n}{L} \right) &=& \frac{L^2}{a^2 \pi^2}\sum_{n'}^{|n'|,|n-n'| \ge n_c} \frac{1}{n'}\frac{1}{n-n'} \ , \label{lastsum}
\end{eqnarray}
where the summation index is integer.

Let us first consider the case $n=0$, where the sum is symmetric to the swap of the sign of $n'$.
Therefore, we can write \rref{lastsum} in the following way:
\begin{eqnarray}
W\left(0 \right) &=& -\frac{2L^2}{a^2\pi^2} \sum_{n'=n_c}^{\infty} \frac{1}{n'^2} \ . \label{lastsum0}
\end{eqnarray}
Using the identity of $\sum_{n'=1}^{\infty}=\pi^2/6$, we can expand \rref{lastsum0} with finite summations, as
\begin{eqnarray*}
W\left(0 \right) &=& -\frac{2L^2}{a^2\pi^2} \left( \frac{\pi^2}{6} -\sum_{n'=1}^{n_c-1} \frac{1}{n'^2}\right) \ .
\end{eqnarray*}
Now let us consider the case $n>0$. If $n<2n_c$  we get the following expression:
\begin{align}
& W\left(\frac{4 \pi n_c}{L} >\frac{2 \pi n}{L}>0\right) = \label{firstpossum} \\ 
&\hspace{0.5cm} \frac{L^2}{a^2\pi^2} {\Bigg (} 
\sum_{n'=-\infty}^{-n_c}\frac{1}{n'}\frac{1}{n-n'}  +  \sum_{n'=n+n_c}^{\infty}\frac{1}{n'}\frac{1}{n-n'}  {\Bigg )} \ , \nonumber
\end{align}
which is supplemented by an additional term, if $n \ge n_c$, as
\begin{align}
&W\left(\frac{2 \pi n}{L} \ge \frac{4 \pi n_c}{L} \right) = 
\frac{L^2}{a^2\pi^2} {\Bigg (} 
\sum\limits_{n'=-\infty}^{-n_c}\frac{1}{n'}\frac{1}{n-n'} \ + 
\label{secondpossum} \\
& \hspace{1.3cm} \ + \  \sum_{n'=n+n_c}^{\infty}\frac{1}{n'}\frac{1}{n-n'}  \
+ \ \sum_{n'=n_c}^{n-n_c}\frac{1}{n'}\frac{1}{n-n'} {\Bigg )} . \nonumber
\end{align}

Let us consider first \rref{firstpossum}, and swap the sign of $n'$ in the first sum and merge all the terms,
where $n'$ is larger than $n+n_c$,
\begin{align}
&W\left(\frac{4 \pi n_c}{L}>\frac{2 \pi n}{L}>0 \right) =  \label{sumres}  \\
& \frac{L^2}{a^2\pi^2} {\Bigg (} 
-\sum_{n'=n_c}^{n+n_c-1}\frac{1}{n'}\frac{1}{n+n'} \ + \nonumber \\
& \hspace{1cm} \ + \
\sum_{n'=n+n_c}^{\infty}
\underbrace{\frac{1}{n'}\left(\frac{1}{n-n'}- \frac{1}{n+n'}  \right)
}_{ \frac{2}{n^2-n'^2}} {\Bigg )} \ . \nonumber
\end{align}
The digamma function $\psi(x)$, can be used to simplify the sum above
by using the following identities:
\begin{align}
&\sum_{n'=a}^\infty \frac{1}{n^2-n'^2} = \frac{\psi(a-n)-\psi(a+n)}{2n} \ , 
\label{digamma1} \\
&\sum_{n'=a}^b \frac{1}{n'}\frac{1}{n+n'} = \label{digamma2} \\ 
& \hspace{1cm} \frac{\psi(1+b)+\psi(a+n)-\psi(1+b+n)-\psi(a)}{n} \ , \nonumber
\end{align}
which can be derived from the series expansion of the digamma function as
\begin{eqnarray*}
\psi(x) &=& -\gamma + \sum_{l=0}^{\infty} \frac{x-1}{(l+1)(l+x)} \ .
\end{eqnarray*}
Using \rrefsa{digamma1} and \rrefsb{digamma2}, \rref{sumres} can be
written in the following form:
\begin{align}
&W\left(\frac{4 \pi n_c}{L}>\frac{2 \pi n}{L}>0 \right) = \label{sumanal} \\
&\hspace{4cm} \frac{2L^2}{a^2 \pi^2 n} \left( \psi(n_c) - \psi(n+n_c) \right) \ .
\nonumber
\end{align}
By using the following property of the digamma function:
\begin{eqnarray*}
\psi(x+1)=\psi(x) + \frac{1}{x} \ ,
\end{eqnarray*}
equation \rrefsb{sumanal} can be written in the following numerically treatable
form:
\begin{eqnarray*}
W\left(\frac{4 \pi n_c}{L}>\frac{2 \pi n}{L}>0\right) &=& -\frac{2L^2}{a^2\pi^2n} \sum_{n'=n_c}^{n+n_c-1} \frac{1}{n'} \ .
\end{eqnarray*}
In the case of $n \ge 2n_c$ and $n<0$, with a similar derivation, we got the following expressions:
\begin{align*}
&W\left(\frac{2 \pi n}{L} \ge \frac{4 \pi n_c}{L}  \right) = -\frac{2L^2}{a^2\pi^2} 
{\Bigg (} \frac{1}{n}\sum_{n'=n_c}^{n+n_c-1} \frac{1}{n'} \ - \\
& \hspace{5cm} - \ \frac{1}{2}\sum_{n'=n_c}^{n-n_c}\frac{1}{n'}\frac{1}{n-n'} {\Bigg )} \ , \\
&W\left(-\frac{4 \pi n_c}{L}<\frac{2 \pi n}{L}<0\right) = \frac{2L^2}{a^2\pi^2n} \sum_{n'=n_c}^{n_c-n-1} \frac{1}{n'} \ , \\
&W\left(\frac{2 \pi n}{L} \le -\frac{4 \pi n_c}{L}\right) = \frac{2L^2}{a^2\pi^2} 
{\Bigg (} \frac{1}{n}\sum_{n'=n_c}^{n_c-n-1} \frac{1}{n'} \ + \\
& \hspace{4.5cm} + \ \frac{1}{2} 
\sum_{n'=n_c-n}^{-n_c}\frac{1}{n'}\frac{1}{n-n'} {\Bigg )} \ .
\end{align*}
The asymptotic expression of $W\left( \frac{2 \pi n}{L} \right)$ for large $n$ can be also given,
\begin{eqnarray}
W\left(\frac{2 \pi n}{L} \right) = \frac{2L^2}{a^2 \pi^2 |n|} \ln(|n|) \ + \ \mathcal{O}\left(\frac{1}{n^2}\right) \ , \label{assymW}
\end{eqnarray}
where we use the following asymptotic expression of the digamma function,
\begin{eqnarray*}
\psi(x)=\ln(x) \ + \ \mathcal{O}\left( \frac{1}{x}\right) \ .
\end{eqnarray*}

\bibliography{transcorrelated,Fermi_gases,Bethe_Peiers,renormalization,scattering,Bose_gas,programs,QMC,Fewfermions,Fewbosons,experiments,2Drealization,book,references,QMC_restofphys,math}

\end{document}